\documentclass[sn-mathphys,Numbered]{sn-jnl}

\usepackage{graphicx}%
\usepackage{multirow}%
\usepackage{amsmath,amssymb,amsfonts}%
\usepackage{amsthm}%
\usepackage{mathrsfs}%
\usepackage[title]{appendix}%
\usepackage{xcolor}%
\usepackage{textcomp}%
\usepackage{manyfoot}%
\usepackage{booktabs}%
\usepackage{algorithm}%
\usepackage{algorithmicx}%
\usepackage{algpseudocode}%
\usepackage{listings}%
\usepackage{bm}
\usepackage{natbib}
\usepackage{longtable}
\usepackage{array}
\usepackage[export]{adjustbox}
\usepackage{rotating}
\usepackage{lineno}

\begin{document}
%\linenumbers
\title[Robust statistical modeling of monthly rainfall: The minimum density power divergence approach]{Robust statistical modeling of monthly rainfall: The minimum density power divergence approach}

\author*[1]{\fnm{Arnab} \sur{Hazra}}\email{ahazra@iitk.ac.in}

\author[2]{\fnm{Abhik} \sur{Ghosh}}\email{abhik.ghosh@isical.ac.in}


\affil*[1]{\orgdiv{Department of Mathematics and Statistics}, \orgname{Indian Institute of Technology Kanpur}, \orgaddress{\city{Kanpur}, \postcode{208016}, \country{India}}}

\affil[2]{\orgdiv{Interdisciplinary Statistical Research Unit}, \orgname{Indian Statistical Institute}, \orgaddress{\city{Kolkata}, \postcode{208016}, \country{India}}}

% \affil[3]{\orgdiv{Department}, \orgname{Organization}, \orgaddress{\street{Street}, \city{City}, \postcode{610101}, \state{State}, \country{Country}}}


\abstract{{\color{black}Statistical modeling of monthly, seasonal, or annual rainfall data is an important research area in meteorology. These models play a crucial role in rainfed agriculture, where a proper assessment of the future availability of rainwater is necessary.} The rainfall amount during {\color{black}a rainy month or a whole rainy season} can take any positive value and some simple (one or two-parameter) probability models supported over the positive real line that are generally used for rainfall modeling are exponential, gamma, Weibull, lognormal, Pearson Type-V/VI, log-logistic, etc., where the unknown model parameters are routinely estimated using the maximum likelihood estimator (MLE). However, the presence of outliers or extreme observations is a common issue in rainfall data and the MLEs being highly sensitive to them often leads to spurious inference. Here, we discuss a robust parameter estimation approach based on the minimum density power divergence estimator (MDPDE). We fit the above four parametric models to the {\color{black} detrended} areally-weighted monthly rainfall data from the 36 meteorological subdivisions of India for the years 1951–2014 and compare the fits based on MLE and the proposed `optimum' MDPDE; the superior performance of MDPDE is showcased for several cases. For all month-subdivision combinations, we discuss the best-fit models and median rainfall amounts.}




\maketitle

% The proportion of the rural population in India is high (68.84\% in 2011; source: \url{http://censusindia. gov. in}) and the main livelihood in the rural areas is agriculture, which contributes 17\% of the country's GDP \citep{arjun2013indian}. 

\section{Introduction}
\label{intro}

{\color{black}Rainfall modeling holds particular significance for rainfed agriculture in India due to the heavy reliance of the country on monsoon rains for crop cultivation; 67\% of the lands are under rainfed agriculture which makes it the largest such extent in the world \citep{venkateswarlu2011rainfed}. The monsoon season significantly influences agricultural outcomes, making accurate rainfall predictions crucial for farmers. In India, the proportion of the rural population is high (68.84\% in 2011; source: \url{http://censusindia. gov. in}) and the main livelihood in the rural areas is agriculture, which contributes 17\% of the country's gross domestic product \citep{arjun2013indian}. In this context, rainfall models play a vital role in crop planning, resource management, and risk mitigation \citep{alam2016statistical}, where farmers can make informed decisions on crop selection, planting times, and water usage based on rainfall forecasts. Given the diverse climatic conditions across different regions of India, precise statistical modeling of rainfall aids in tailoring agricultural practices to local needs, helping communities adapt to the challenges posed by variable and changing precipitation patterns. Moreover, such models secondarily contribute to the effective implementation of government policies and initiatives aimed at supporting sustainable and resilient rainfed agriculture in the country.}

Statistical modeling of monthly, seasonal, or annual total rainfall has been an important research area in meteorology over the decades. {\color{black}Considering the wet months, i.e., the months with the amount of monthly total rainfall being nonzero, usually, the histograms appear to be positively skewed.} Hence, the probability distributions justified for this purpose are right-skewed and defined over the whole positive real line; some examples are exponential \citep{todorovic1975stochastic,burgueno2005statistical,burgueno2010statistical,hazra2018bayesian}, gamma \citep{barger1949evaluation, mooley1968application,husak2007use,krishnamoorthy2014small,martinez2019precipitation}, 
log-normal \citep{kwaku2007characterization, mandal2015estimation, adham2016water}, Weibull \citep{duan1995comparison,burgueno2005statistical,lana2017rainfall}, Pearson Type-V/VI \citep{hanson2008probability, khudri2013determination, mayooran2014statistical}, log-logistic  \citep{fitzgerald2005analysis, sharda2005modelling}, generalized exponential \citep{madi2007bayesian, kazmierczak2015suitability,hazra2022minimum}, and generalized gamma \cite{sharma2010use, mandal2015estimation, mamoon2017selection} distributions. Out of several possible choices, the first four have been used predominantly for rainfall modeling, and hence, we also concentrate only on those four distributions in the main article. Henceforth, we refer to them as `rainfall models' or simply as RM. Other than the above-mentioned probability models, several researchers have proposed using two-parameter exponential, three-parameter gamma, three-parameter lognormal, and three-parameter Weibull distributions \citep{sharma2010use, maposa2014investigating, mandal2015estimation, kumar2017statistical} in the context of rainfall modeling as well as in other related hydrology literature. These models include an additional location parameter $\mu_0$, along with those in RM, controlling the support of the probability distributions which is $(\mu_0, \infty)$. This is less intuitive as the rainfall amount in a wet month can take any positive value. If the estimated location parameter is $\hat{\mu}_0 = 2$ mm (say), this infers that there is no chance of observing less than 2 mm of rain in the future months. Besides, there must not be a positive probability of observing a negative amount of rainfall. However, this is the case when the estimated location parameter is negative. Such models with a location parameter are thus excluded from consideration in our study. On the other hand, some max-stable distributions like the generalized extreme value distribution and the generalized Pareto distribution and their generalizations like a four-parameter kappa distribution have also been used for rainfall modeling and in hydrology \citep{mandal2015estimation, mamoon2017selection, kumar2017statistical}. However, following extreme value theory, these models are only applicable while modeling annual/monthly maxima of daily rainfall amounts or while modeling only the tail distribution of the rainfall amounts. Thus, these models are not theoretically justified for monthly or annual total rainfall amounts, despite being used in hydrology. We do not consider them in our study as our interest lies in modeling monthly totals, neither block maxima nor threshold exceedances. Mostly, researchers choose one such model and analyze the data based on it. However, different goodness-of-fit tests have also been used in the literature for data-based selection of an appropriate model; these include the chi-square test \citep{barger1949evaluation, mooley1973gamma, kwaku2007characterization}, Kolmogorov-Smirnov test \citep{sharma2010use, hazra2014modelling, al2014frequency}, Anderson-Darling test \citep{sharma2010use}, variance ratio test \citep{mooley1973gamma, hazra2017note} and the Akaike information criterion (AIC) used by \cite{villarini2012development}.

% , strupczewski2007robustness

The maximum likelihood estimation (MLE) is the most widely used parameter estimation procedure in the meteorology literature. It has several attractive asymptotic properties like consistency and full asymptotic efficiency; it achieves the Cram\'{e}r-Rao lower bound when the sample size tends to infinity. However, the MLE is highly sensitive to outliers and gets strongly affected even in the presence of a single outlying observation \citep{strupczewski2005robustness,neykov2007robust,strupczewski2007robustness}. For robust inference, the $L$-moments estimation (LME, henceforth) proposed by \cite{hosking1990moments} is widely used in hydrology. In LME, instead of obtaining the parameter estimates by equating the raw/central population moments and the sample moments as in the method of moments estimation, one equates the population moments and the sample moments of some linear combinations of order statistics.

{\color{black} As an alternative to the existing robust parameter estimation procedures, \cite{basu1998robust} propose the minimum density power divergence estimation (MDPDE). Here, we obtain the estimates by minimizing a suitable density-based divergence measure, known as the density power divergence (DPD), over the parameter space.} The DPD family is indexed by a tuning parameter $\alpha\geq 0$, and the family reduces to the well-known Kullback-Leibler divergence at $\alpha=0$. Thus, the MDPDE at $\alpha=0$ is the same as MLE, and it provides a robust generalization of the MLE at $\alpha>0$. A significant advantage of the MDPDE approach is that it does not need any nonparametric smoothing for the density estimation like other minimum divergence approaches \citep{beran1977minimum,basu1994minimum}; as a result, MDPDE is comparatively easy to implement in practice \citep{seo2017extreme}. As a result, the MDPDE has been widely used in several areas of scientific research like pollution monitoring \citep{gajewski2004correspondence},  Gene Ontology \citep{yuan2008partial}, and meteorology \citep{seo2017extreme}. While \cite{seo2017extreme} used MDPDE for modeling extreme rainfall (annual maximum daily rainfall, precipitation above some high threshold), as of authors' knowledge, it has not yet been used in classical statistical meteorology literature (e.g., modeling monthly or annual rainfall).  %Because of using order statistics, this estimation approach is considered to be more robust.}

% , namely exponential, gamma, lognormal, and Weibull (we refer to them as ``rainfall model"s or simply as RMs),

In this paper, we analyze {\color{black} detrended} areally-weighted monthly rainfall data from the 36 meteorological subdivisions of India during 1951--2014 because of its importance in rainfed agriculture across the country. {\color{black}Due to the possible presence of outliers, we detrend the data using $L_1$ regression and analyze the residuals subsequently.} Implementing the widely-used Adjusted-Boxplot approach of \cite{hubert2008adjusted}, we identify that the proportions of outliers are high in the {\color{black} residuals}, particularly during the monsoon months. We fit the four probability distributions in RM and estimate the model parameters using the proposed MDPDE for all subdivision-month combinations. For each value of the tuning parameter $\alpha$, the MDPDE returns a separate estimate of the model parameters; the asymptotic relative efficiency of the MDPDEs decreases with increasing values of $\alpha$ only moderately. However, such loss in efficiency is observed not to be quite significant compared to the huge gain in terms of robustness, as illustrated through the classical influence function analysis, for all four members of RM. We study how the fitted RM members vary over different choices of the tuning parameter $\alpha$ and describe an optimal data-driven selection of $\alpha$ by minimizing the empirical Wasserstein distance \citep{vallender1974calculation} following a leave-one-out cross-validation procedure mentioned in \cite{fujisawa2006robust}. Instead of the non-robust AIC for model selection under MLE, we use robust information criterion (RIC) based on the MDPDEs to select the best-fit model among the four RM members considered. For each RM, we present results at four subdivision-month combinations to illustrate the advantage of the MDPDE-based approach over the non-robust MLE and robust LME-based analyses. We finally provide the best-fit models and the median rainfall amounts based on the `optimum' MDPDE estimates for all month-subdivision combinations.

The rest of the paper is structured as follows. In Section \ref{data}, we discuss the Indian rainfall dataset used here, along with some {\color{black} data preprocessing steps and} preliminary analyses. The statistical methodology and illustrations of their properties for the four RMs are presented in Section \ref{method}. We discuss the results in Section \ref{results}, and the paper ends with some concluding remarks in Section \ref{conclusion}.

% Anderson-Darling (AD)

\section{Rainfall data}
\label{data}

\begin{figure}[t]
    \centering
    \includegraphics[width = 0.45\linewidth]{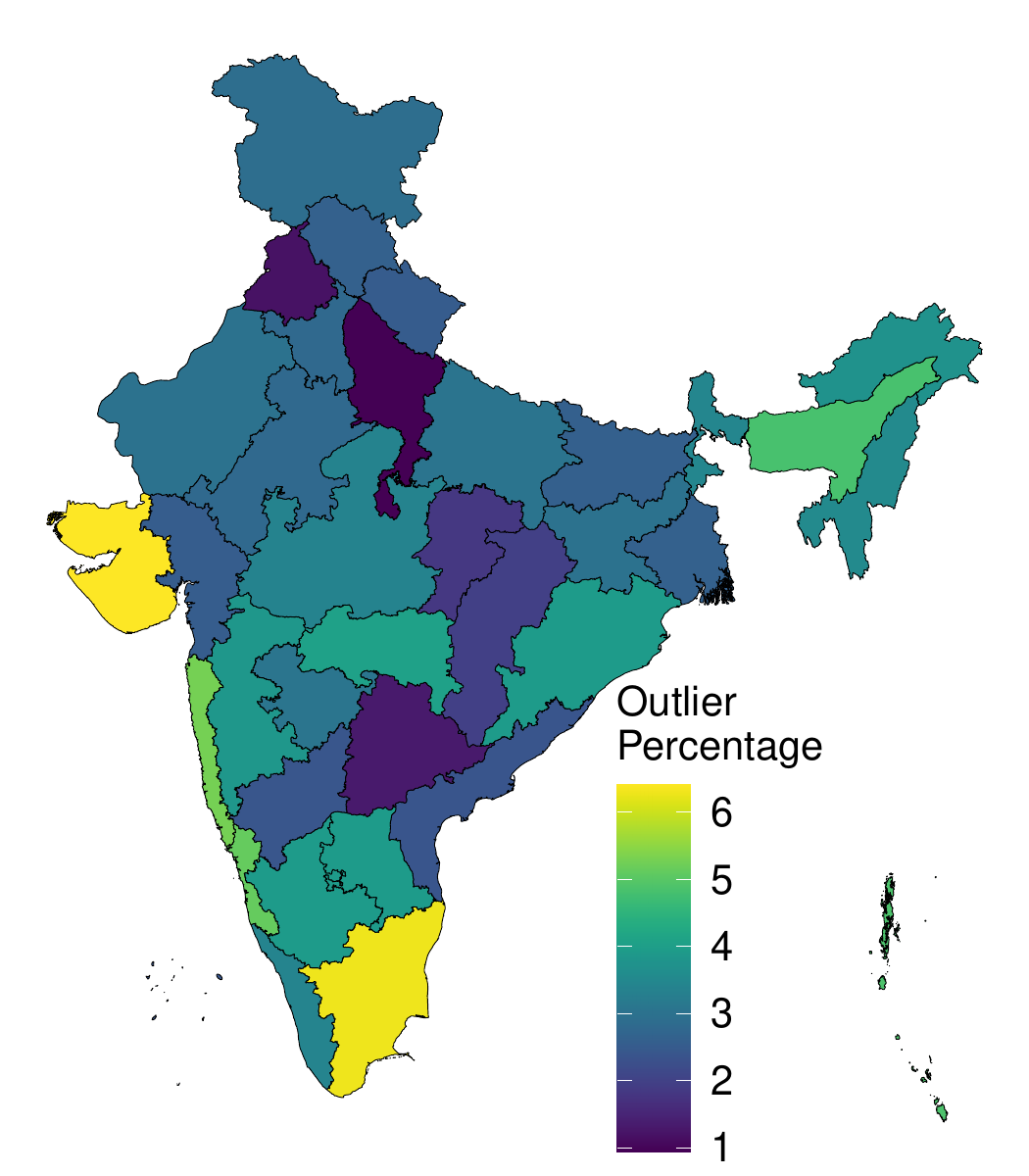}
    \includegraphics[width = 0.45\linewidth]{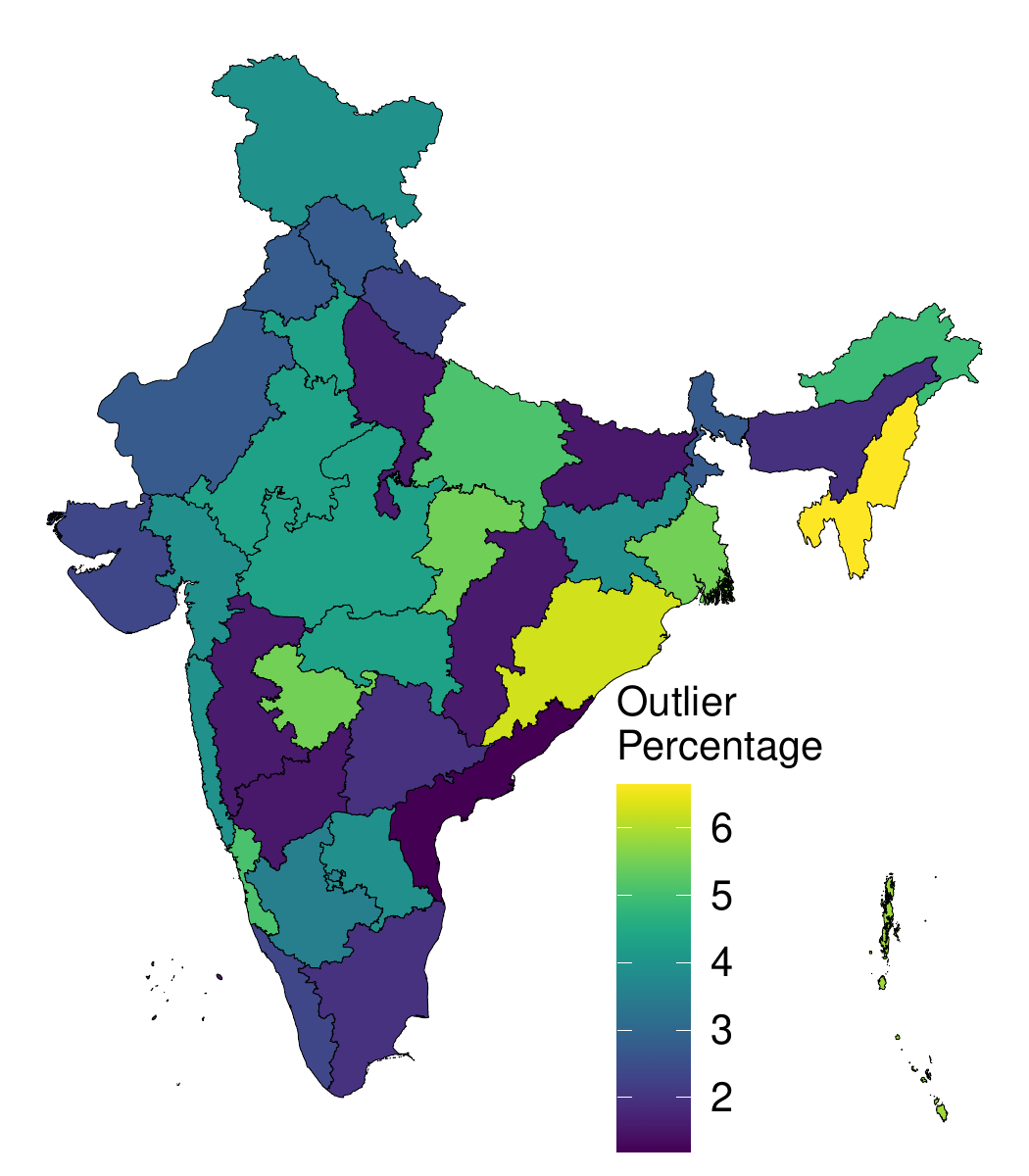}
    \caption{{\color{black}Monthly average percentage of outliers, for all months (left) and only the monsoon months of June through September (right), present in detrended rainfall data for 36 meteorological subdivisions of India, obtained using the Adjusted-Boxplot method of \cite{hubert2008adjusted}.}}
    \label{fig:outlierprop}
\end{figure}

{\color{black} Unlike the administrative state-wise boundaries, from the perspective of meteorological homogeneity, India is divided into 36 meteorological subdivisions \cite{guhathakurta2008trends}; the geographical boundaries of these regions are presented in Figure \ref{fig:outlierprop}. We obtain areally-weighted monthly rainfall (in mm) data for these meteorological subdivisions from the Open Government Data (OGD) Platform, India (\url{https://data.gov.in}) and shapefile of the boundaries from \url{https://data.gov.in}. For preparing the dataset, India Meteorological Department (IMD) performed rainfall monitoring in 641 districts across India and \cite{guhathakurta2008trends} computed monthly rainfall amounts for these districts by averaging the rainfall amounts at available stations within the respective district for each month. Further, to obtain subdivision-wise rainfall data from the district-wise rainfall data, an area-weighted averaging method was followed. A more detailed report on the dataset preparation is in \cite{guhathakurta2011new}. In this paper, we analyze the monthly rainfall data for the 36 meteorological subdivisions of India available over the years 1951--2014.}

% More details on year-wise numbers of stations used for obtaining the district-level data are in \cite{guhathakurta2011new}. 

{\color{black} Due to the long observational period of 64 years, specifically covering the period that faced global warming, the presence of trends is likely in this dataset. Due to the possible presence of outliers in the dataset, we use the robust nonparametric Cox and Stuart trend test \citep{cox1955some} using the function \texttt{cs.test} implemented in the \texttt{R} package \texttt{trends} \citep{pohlert2023trend}. Out of the total 432 subdivision-month combinations, trends are significant in 35 and 17 cases at the significance levels 5\% and 1\%, respectively. As a robust alternative to simple linear regression, we perform $L_1$ regression to estimate the trends in (log-transformed) nonzero rainfall data and obtain the residuals; to avoid confusion, we henceforth use the term `detrended data' to denote the (exponentiated) residuals, i.e., our pre-processing step removes the inhomogeneity in the scale parameters. The Cox and Stuart trend test ensures that the detrended data do not have any significant trend for any subdivision-month combination and we go ahead with modeling the detrended data with the commonly used IID setup for MDPDE.}

% Ignoring these few cases, for each subdivision-month combination, we assume that the monthly rainfall amounts across the years are independent and identically distributed. These assumptions are common for modeling rainfall data in meteorological literature \citep[see, e.g.,][]{mooley1973gamma,sharda2005modelling,hazra2017note}.

% Due to the possible presence of outliers in the dataset, a
% Given the MDPDE methodology mainly available for an IID data setup, we first detrend the data by fitting a linear trend model. 
% to the log-transformed nonzero rainfall data
% the $p$-value of the related hypothesis testing problem

\begin{sidewaystable}
\centering
\caption{{\color{black}Percentages of outliers present in the detrended data for each month at the 36 meteorological subdivisions of India, identified based on Adjusted-Boxplot approach of \cite{hubert2008adjusted}}}
\label{table1}
{\color{black}
%\begin{adjustbox}{angle=90}
%\Rotatebox{90}{
\begin{tabular}{lrrrrrrrrrrrr}
  \hline
Meteorological subdivision & Jan & Feb & Mar & Apr & May & Jun & Jul & Aug & Sep & Oct & Nov & Dec \\ 
  \hline
Andaman and Nicobar islands & 1.6 & 5.1 & 0.0 & 7.9 & 4.7 & 4.7 & 3.1 & 12.5 & 3.1 & 1.6 & 1.6 & 12.5 \\ 
  Arunachal Pradesh & 0.0 & 0.0 & 0.0 & 0.0 & 11.5 & 8.2 & 4.9 & 4.9 & 1.6 & 1.7 & 1.7 & 10.3 \\ 
  Assam and Meghalaya & 1.6 & 4.7 & 10.9 & 15.6 & 4.7 & 0.0 & 1.6 & 6.2 & 0.0 & 3.1 & 4.7 & 4.7 \\ 
  Nagaland,Manipur,Mizoram and Tripura & 0.0 & 4.8 & 6.3 & 1.6 & 3.1 & 12.5 & 7.8 & 3.1 & 3.1 & 0.0 & 0.0 & 0.0 \\ 
  Sub-Himalayan West Bengal and Sikkim & 3.2 & 0.0 & 1.6 & 10.9 & 0.0 & 3.1 & 4.7 & 0.0 & 3.1 & 0.0 & 6.3 & 8.5 \\ 
  Gangatic West Bengal & 0.0 & 0.0 & 0.0 & 3.1 & 1.6 & 1.6 & 10.9 & 1.6 & 7.8 & 4.7 & 0.0 & 0.0 \\ 
  Orissa & 0.0 & 1.6 & 7.8 & 4.7 & 6.2 & 9.4 & 6.2 & 7.8 & 1.6 & 1.6 & 0.0 & 0.0 \\ 
  Jharkhand & 1.6 & 1.6 & 0.0 & 3.1 & 10.9 & 1.6 & 6.2 & 4.7 & 3.1 & 3.1 & 0.0 & 0.0 \\ 
  Bihar & 0.0 & 0.0 & 0.0 & 1.6 & 0.0 & 0.0 & 0.0 & 6.2 & 0.0 & 17.2 & 6.0 & 0.0 \\ 
  East Uttar Pradesh & 1.6 & 11.3 & 0.0 & 1.7 & 0.0 & 1.6 & 1.6 & 3.1 & 14.1 & 1.6 & 0.0 & 0.0 \\ 
  West Uttar Pradesh & 0.0 & 0.0 & 3.4 & 1.6 & 0.0 & 4.7 & 1.6 & 0.0 & 0.0 & 0.0 & 0.0 & 0.0 \\ 
  Uttarakhand & 0.0 & 3.1 & 1.6 & 1.6 & 1.6 & 1.6 & 1.6 & 3.1 & 3.1 & 12.7 & 0.0 & 0.0 \\ 
  Haryana,Chandigarh and Delhi & 3.2 & 1.6 & 0.0 & 1.6 & 0.0 & 6.2 & 4.7 & 1.6 & 4.7 & 3.8 & 4.2 & 1.9 \\ 
  Punjab & 0.0 & 0.0 & 0.0 & 1.6 & 0.0 & 1.6 & 3.1 & 1.6 & 4.7 & 1.6 & 0.0 & 0.0 \\ 
  Himachal Pradesh & 0.0 & 4.7 & 6.2 & 0.0 & 4.7 & 7.8 & 0.0 & 1.6 & 1.6 & 3.2 & 0.0 & 1.6 \\ 
  Jammu and Kashmir & 4.8 & 3.1 & 4.7 & 1.6 & 1.6 & 0.0 & 6.2 & 1.6 & 7.9 & 0.0 & 0.0 & 3.2 \\ 
  West Rajasthan & 0.0 & 0.0 & 1.8 & 5.1 & 0.0 & 7.8 & 0.0 & 3.1 & 0.0 & 3.4 & 2.6 & 12.2 \\ 
  East Rajasthan & 1.8 & 1.9 & 1.8 & 1.7 & 0.0 & 0.0 & 3.1 & 10.9 & 3.1 & 0.0 & 2.6 & 6.4 \\ 
  West Madhya Pradesh & 13.6 & 0.0 & 0.0 & 0.0 & 1.6 & 9.4 & 4.7 & 3.1 & 0.0 & 0.0 & 0.0 & 8.3 \\ 
  East Madhya Pradesh & 0.0 & 0.0 & 0.0 & 0.0 & 0.0 & 7.8 & 3.1 & 4.7 & 6.2 & 0.0 & 0.0 & 0.0 \\ 
  Gujarat Reg. Daman and Diu and Nagar Havelli & 0.0 & 0.0 & 7.4 & 0.0 & 0.0 & 4.7 & 6.2 & 4.7 & 0.0 & 0.0 & 7.3 & 0.0 \\ 
  Saurashtra,Kutch and Diu & 5.9 & 0.0 & 16.0 & 14.3 & 2.3 & 0.0 & 7.8 & 1.6 & 0.0 & 0.0 & 2.8 & 26.1 \\ 
  Konkan and Goa & 20.0 & 0.0 & 8.7 & 0.0 & 3.2 & 12.5 & 1.6 & 1.6 & 0.0 & 0.0 & 10.0 & 5.6 \\ 
  Madhya Maharashtra & 2.3 & 0.0 & 5.6 & 0.0 & 0.0 & 1.6 & 0.0 & 4.7 & 0.0 & 17.2 & 5.4 & 9.3 \\ 
  Marathawada & 0.0 & 0.0 & 5.5 & 0.0 & 1.6 & 1.6 & 14.1 & 6.2 & 0.0 & 4.8 & 0.0 & 2.9 \\ 
  Vidarbha & 0.0 & 3.6 & 0.0 & 14.3 & 1.6 & 0.0 & 0.0 & 1.6 & 15.6 & 0.0 & 11.8 & 0.0 \\ 
  Chattisgarh & 0.0 & 1.7 & 0.0 & 7.9 & 1.6 & 1.6 & 3.1 & 1.6 & 0.0 & 6.2 & 0.0 & 0.0 \\ 
  Coastal Andhra Pradesh & 1.7 & 0.0 & 1.6 & 1.6 & 17.2 & 1.6 & 1.6 & 1.6 & 0.0 & 0.0 & 1.6 & 0.0 \\ 
  Telangana & 0.0 & 0.0 & 0.0 & 4.7 & 3.1 & 4.7 & 1.6 & 1.6 & 0.0 & 0.0 & 0.0 & 0.0 \\ 
  Rayalaseema & 0.0 & 4.5 & 2.0 & 12.5 & 4.7 & 7.8 & 4.7 & 3.1 & 0.0 & 6.2 & 1.6 & 0.0 \\ 
  Tamil Nadu and Pondicherry & 9.4 & 0.0 & 9.5 & 17.2 & 9.4 & 1.6 & 4.7 & 0.0 & 1.6 & 12.5 & 0.0 & 9.4 \\ 
  Coastal Karnataka & 3.1 & 10.7 & 1.9 & 0.0 & 17.2 & 7.8 & 7.8 & 3.1 & 1.6 & 1.6 & 4.8 & 1.8 \\ 
  North Interior Karnataka & 2.8 & 8.8 & 1.6 & 1.6 & 0.0 & 3.1 & 3.1 & 0.0 & 0.0 & 7.8 & 0.0 & 0.0 \\ 
  South Interior Karnataka & 0.0 & 0.0 & 11.1 & 3.1 & 1.6 & 0.0 & 14.1 & 0.0 & 0.0 & 1.6 & 15.6 & 0.0 \\ 
  Kerala & 1.6 & 3.2 & 14.1 & 4.7 & 7.8 & 4.7 & 1.6 & 3.1 & 0.0 & 0.0 & 0.0 & 0.0 \\ 
  Lakshadweep & 3.2 & 0.0 & 0.0 & 16.1 & 1.6 & 6.2 & 0.0 & 0.0 & 0.0 & 1.6 & 0.0 & 0.0 \\ 
   \hline
\end{tabular}
}%\end{adjustbox}
%}
\end{sidewaystable}

% Similar to \cite{seo2017extreme}, we treat an observation to be an outlier if it is greater than the 1.5 interquartile range above the third quartile or lower than the 1.5 interquartile range below the first quartile.

We identify outliers present in the {\color{black} detrended} data using the widely-used Adjusted-Boxplot method \cite{hubert2008adjusted} implemented in the \texttt{R} package \texttt{univOutl} \citep{univOutl}. In the context of robust estimation for extreme rainfall modeling, \cite{seo2017extreme} treat an observation as an outlier if it is greater than the 1.5 interquartile range above the third quartile or lower than the 1.5 interquartile range below the first quartile. However, rainfall data often exhibit skewness, and the Adjusted-Boxplot method is specifically built for skewed distributions, where the main idea is to use a robust skewness measure that is not sensitive to the outliers while determining the whisker length. For each subdivision-month combination, we calculate the proportions of outliers. {\color{black} The highest proportion is at the subdivision Saurashtra, Kutch, and Diu for December (26.1\%), whereas there is no outlier for 162 subdivision-month combinations. After averaging across the subdivisions, the proportion is highest for April (4.53\%) and lowest for February (2.11\%). The subdivision-wise percentages of outliers, after averaging across all twelve months and also averaging over only the monsoon months of June through September, are presented in Figure \ref{fig:outlierprop}. In the first case, the highest proportion is for the subdivision Saurashtra, Kutch, and Diu (6.4\%), and it is the lowest for the West Uttar Pradesh subdivision (0.94\%). In the second case, the highest proportion is for the subdivision Nagaland, Manipur, Mizoram, and Tripura (6.62\%), and it is the lowest for the Coastal Andhra Pradesh subdivision (1.2\%). In the first case, out of 36 regions, 31 regions (except West Uttar Pradesh, Punjab, East Madhya Pradesh, Chattisgarh, and Telangana) have more than 2\% outliers. In the second case, 26 regions (except Assam and Meghalaya, Bihar, West Uttar Pradesh, Madhya Maharashtra, Chattisgarh, Coastal Andhra Pradesh, Telangana, Tamil Nadu and Pondicherry, North Interior Karnataka, and Lakshadweep) have more than 2\% outliers. These findings indicate the importance of a robust inference for a general monthly analysis as well as an analysis from an agro-meteorology perspective. Here the proportion of outliers is unreliably high for certain subdivision-month combinations due to the availability of a very limited number of observations. For example, 26.1\% outliers occur in the case of the subdivision Saurashtra, Kutch, and Diu for December, where only 23 out of 64 years of the observation period received nonzero rainfall. As a result, the estimated proportion of outliers based on the popular Adjusted-Boxplot method might be unreliable. However, this issue is mostly observed for certain subdivision-month pairs that are less relevant from an agro-meteorology perspective. Despite this issue, a large number of agro-meteorologically important subdivision-month pairs exhibit a reasonably high proportion of outliers which motivates us to consider a robust modeling procedure.}

\section{Statistical methodology}
\label{method}

\subsection{The minimum density power divergence estimator}

In this subsection, we summarize the proposed minimum density power divergence estimator (MDPDE);
see \cite{basu1998robust, basu2019statistical} for more details. 

Suppose we have independent and identically distributed (IID) observations $X_1, \ldots, X_n$ 
from a population having true distribution function $G$ and density function $g$.
We want to model it by a parametric family of distribution functions $\lbrace F_\theta \rbrace$ 
having densities $\lbrace f_\theta \rbrace$, indexed by some unknown $p$-dimensional parameter $\theta \in \Theta$, the parameter space. Note that the density functions exist for our four RMs. We need to estimate the unknown model parameter $\theta$ based on the observed data for further inference.

In the common maximum likelihood estimation, we calculate the likelihood function $L(\theta) = \prod_{i=1}^{n} f_\theta(X_i)$ and maximize it over the parameter space $\Theta$ to get the MLE, i.e., $\widehat{\theta}_{\textrm{MLE}} = \arg \max\limits_{\theta \in \Theta} L(\theta) 
= \arg \min\limits_{\theta \in \Theta} \sum_{i=1}^{n} -\log[f_\theta(X_i)]$. In the case of the RMs, a unique (but non-robust) estimate can be obtained.

In an alternative minimum divergence approach, one may consider an appropriate divergence measure between 
the true data-generating density (estimated from the observed data) and the parametric model density and minimize this measure of discrepancy with respect to the underlying model parameter to obtain the corresponding minimum divergence estimate. The MLE can also be considered a minimum divergence estimator associated with the Kullback-Leibler divergence. However, an appropriate choice of the divergence measure is important when our goal lies in robust parametric inference. Among many such available divergences, as mentioned before, here we consider particularly the DPD measure proposed by \cite{basu1998robust}. For a tuning parameter $\alpha\geq 0$, the density power divergence $d_\alpha$ between two densities $f$ and $g$ is defined as
\begin{equation}\label{EQ:dpd}
    d_\alpha(g,f) = \displaystyle \left\{\begin{array}{ll}
    \displaystyle \int  \left[f^{1+\alpha}(x) - \left(1 + \frac{1}{\alpha}\right)  f^\alpha(x) g(x) + 
\frac{1}{\alpha} g^{1+\alpha}(x)\right] dx, & {\rm ~for} ~\alpha > 0,\\
	\displaystyle \lim_{\alpha\rightarrow 0}d_\alpha(g, f) = \int g(x) \left[\log g(x) - \log f(x) \right] dx, & {\rm ~for} ~\alpha = 0.  
\end{array}\right.
\end{equation}
Here, $d_0(g, f)$ is the Kullback-Leibler (KL) divergence. For the case of parametric estimation, we consider the model density $f_\theta$ in place of the density $f$ in Equation (\ref{EQ:dpd}),
whereas $g$ denotes the true density. Then, we can define the minimum DPD functional $T_\alpha(\cdot)$ by \citep{basu1998robust}
\begin{equation}\label{EQ:dpd2}
d_\alpha(g, f_{T_\alpha(G)}) = \min_{\theta \in \Theta} d_\alpha(g, f_\theta),
\end{equation}
whenever the minimum is attained. Thus, $T_\alpha(G)$ represents the best fitting parameter value under the true distribution $G$. In practice, however, the true density $g$ is unknown, and hence the minimizer of $d_\alpha(g, f_\theta)$ cannot be obtained directly; alternatively, we use an estimate of $g$. A major advantage of the particular DPD family over other robust divergence measures is that we can avoid the nonparametric smoothing (and associated numerical complications) for the purpose of estimating density $g$. To see this, we rewrite the DPD measure as 
\begin{equation}\label{EQ:dpd3}
    d_\alpha(g,f_\theta) = \displaystyle \left\{\begin{array}{ll}
    \displaystyle \int  f_\theta^{1+\alpha}(x) dx - \left(1 + \frac{1}{\alpha}\right) E \left[f_\theta^\alpha(X) \right] + 
\frac{1}{\alpha} E\left[ g^{\alpha}(X)\right] & {\rm for} ~\alpha > 0,\\
	\displaystyle E\left[\log g(X)\right] - E\left[\log f_\theta(X)\right] & {\rm for} ~\alpha = 0.  
\end{array}\right.
\end{equation}
where $E\left[ \cdot \right]$ denotes the expectation of its argument with respect to the true density $g$. Note that the terms $E\left[g^\alpha(X) \right]$ and $E\left[\log g(X)\right]$ do not depend on $\theta$ and hence they can be ignored while performing optimization with respect to $\theta$; and the other two expectations in (\ref{EQ:dpd3}) can directly be estimated through empirical means avoiding the direct non-parametric estimation of $g$. 
Therefore, the minimum DPD estimator (MDPDE) is finally defined as
\begin{equation}\label{EQ:dpd4}
\widehat{\theta}_\alpha = \arg \min_{\theta \in \Theta} H_{\alpha, n}(\theta),
\end{equation}
where $H_{\alpha, n}(\theta) = \frac{1}{n} \sum\limits_{i=1}^n V_\alpha(\theta; X_i)$
% is the empirical version of $d_\alpha(g, f_\theta)$ and 
with
\begin{equation}\label{EQ:dpd5}
   V_\alpha(\theta; x)  = \displaystyle \left\{\begin{array}{ll}
    \displaystyle {\color{black}\int  f_\theta^{1+\alpha}(z) dz} - \left(1 + \frac{1}{\alpha}\right) f_\theta^\alpha(x) ,  & {\rm ~~for} ~\alpha > 0,\\
	\displaystyle - \log f_\theta(x), & {\rm ~~for} ~\alpha = 0.  
\end{array}\right.
\end{equation}
%The third term $1/\alpha$ in case of $\alpha > 0$ ensures $V_0(\theta; x) = \lim_{\alpha\rightarrow 0}V_\alpha(\theta; x)$. 
Further, at $\alpha = 0$, $\widehat{\theta}_0 = \arg \min_{\theta \in \Theta} \frac{1}{n} \sum\limits_{i=1}^n [- \log f_\theta(X_i)]$
is clearly the MLE, by definition. However, for any $\alpha\geq 0$, we obtain unbiased estimating equations through the differentiation of $H_{\alpha, n}(\theta)$ in (\ref{EQ:dpd4}) and they are given by 
\begin{equation}\label{EQ:dpd6}
U_n(\theta) \equiv \frac{1}{n} \sum_{i=1}^{n} u_\theta (X_i) f_\theta^\alpha(X_i) - \int u_\theta (x) f_\theta^{1 + \alpha} (x) dx = 0,
\end{equation}
where $u_\theta (x)  = \delta \log f_\theta(x) / \delta \theta$ is the score function. 
Once again at $\alpha=0$, (\ref{EQ:dpd6}) reduces to the usual score equation leading to the MLE.
But, for $\alpha>0$, the MDPDEs provide a weighted score equation (suitably adjusted for unbiasedness)
with weights $f_\theta^\alpha(X_i)$ for $X_i$. These weights would be small for outlying observations (with respect to the model family) and thus are expected to produce robust estimates by downweighting the effects of outliers. 

\subsection{The MDPDE for the rainfall models}
\label{MDPDE for rainfall models}

Considering the intricacy of the mathematical details, 
we illustrate the basic steps to obtain the MDPDEs for our four RMs; 
%exponential distribution in details and provide brief overview for the rest of the  RMs in this subsection. We skip 
the rigorous differentiation and integration steps are omitted for brevity.

\subsubsection{Exponential distribution}\label{subsec:exponential}

Consider the family of one-parameter exponential distributions having distribution function $F_\lambda(x) = 1 - \exp{(-\lambda x)}$, 
and the associated density function $f_\lambda(x) =  \lambda \exp{(-\lambda x)}$, for $x>0$, 
where $\lambda>0$ denotes the rate parameter. 
Straightforward calculations of the terms $V_\alpha(\theta; x)$ in (\ref{EQ:dpd5}) show that 
\begin{equation}\label{EQ:dpd5:exp}
   V_\alpha(\lambda; x)  = \displaystyle \left\{\begin{array}{ll}
    \displaystyle \frac{\lambda^\alpha}{1 + \alpha}  - \left(1 + \frac{1}{\alpha}\right) \lambda^\alpha \exp{(-\alpha \lambda x)},       & {\rm ~for} ~\alpha > 0,\\
	\displaystyle \lambda x- \log(\lambda), & {\rm ~for} ~\alpha = 0.  
\end{array}\right.
\end{equation}

While calculating the estimating equation from (\ref{EQ:dpd4}), the score function is $u_\lambda (x)  = \lambda^{-1} - x$. 
Plugging this, we get	 
\begin{equation}\label{EQ:dpd6:exp}
U_n(\lambda) \equiv \frac{1}{n} \sum_{i=1}^{n} \left( \frac{1}{\lambda} - X_i \right) \lambda^\alpha \exp{(-\alpha \lambda X_i)} - \frac{\alpha \lambda^{\alpha - 1}}{(1 + \alpha)^2} = 0.
\end{equation}
The MDPDE estimate of $\lambda$ is then obtained by solving (\ref{EQ:dpd6:exp}). A closed-form expression of $\widehat\lambda_\alpha$ does not exist for $\alpha>0$ and hence, we compute them by solving (\ref{EQ:dpd6:exp}) numerically.

\subsubsection{Gamma distribution}\label{subsec:gamma}

Here, we assume that we have IID observations $X_1, \ldots, X_n$ from the two-parameter gamma distribution family. 
The corresponding distribution function $F_{(a,b)}(x)$ does not have a closed-form expression, but the density function has the form
$f_{(a,b)}(x) =  \frac{b^a}{\Gamma(a)} x^{a-1} \exp{(-b x)}$, for $x>0$, where $a$ and $b$ denote the shape and rate parameters, respectively. 
Straightforward calculations of $V_\alpha(\theta; x)$ from (\ref{EQ:dpd5}) yields 
\begin{equation}\label{EQ:dpd5:gamma}
   V_\alpha(a, b; x)  = \displaystyle \left\{\begin{array}{ll}
    \displaystyle \frac{\Gamma\left( (a-1) (1 + \alpha) + 1 \right) b^\alpha}{\Gamma(a)^{\alpha+1} (1 + \alpha)^{(a-1) (1 + \alpha) + 1}}
    \\
   \displaystyle  ~~~~ - {\color{black}\left(1 + \frac{1}{\alpha}\right) \frac{b^{\alpha a}}{\Gamma(a)^\alpha}} x^{\alpha(a-1)} \exp{(-\alpha b x)},  & {\rm ~for} ~\alpha > 0,\\
	\displaystyle \log \Gamma(a) - a \log(b) + bx - (a-1) \log(x), & {\rm ~for} ~\alpha = 0.  
\end{array}\right.
\end{equation}

Thus, we calculate $H_{\alpha, n}(a, b) = \frac{1}{n} \sum_{i=1}^n V_\alpha(a, b; X_i)$ by plugging in the observations in (\ref{EQ:dpd5:gamma}) and obtain the MDPDE of $a$ and $b$ by minimizing $H_{\alpha, n}(a, b)$ numerically.

\subsubsection{Lognormal distribution}\label{subsec:lognormal}

We next consider the two-parameter lognormal distribution family having a distribution function 
$F_{(\mu,\sigma)}(x) = \Phi \left( \frac{\log(x) - \mu}{\sigma} \right)$, 
where $\Phi$ is the distribution function of a standard normal distribution and 
$\mu$ and $\sigma$ denote the mean and standard deviation parameters in the log scale, respectively. 
The corresponding density function is given by 
$f_{(\mu,\sigma)}(x) = \frac{1}{\sqrt{2 \pi} \sigma x} \exp{\left( - \frac{(\log(x) - \mu)^2}{2\sigma^2} \right)}$, 
for $x > 0$. Once again, straightforward calculations show that 
\begin{equation}\label{EQ:dpd5:lnorm}
   V_\alpha(\mu, \sigma; x)  = \displaystyle \left\{\begin{array}{lll}
    \displaystyle \frac{1}{\sqrt{\alpha + 1} (\sqrt{2 \pi} \sigma)^\alpha} \exp{\left( -\alpha \mu + \frac{\alpha^2 \sigma^2}{2(\alpha + 1)} \right)} 
   \\
  \displaystyle ~~~~ - \left(1 + \frac{1}{\alpha}\right) \frac{1}{(2 \pi)^{\alpha / 2} \sigma^\alpha x^\alpha} \exp{\left( - \alpha \frac{(\log(x) - \mu)^2}{2\sigma^2} \right)},  & {\rm ~for} ~\alpha > 0,\\
	\displaystyle \log \left(\sqrt{2 \pi} \sigma x\right) + \frac{(\log(x) - \mu)^2}{2\sigma^2}, & {\rm ~for} ~\alpha = 0.  
\end{array}\right.
\end{equation}

Plugging in $X_1, \ldots, X_n$ in (\ref{EQ:dpd5:lnorm}), we again compute the objective function 
$H_{\alpha, n}(\mu, \sigma) = \frac{1}{n} \sum_{i=1}^n V_\alpha(\mu, \sigma; X_i)$, 
and then obtain the MDPDE of $\mu$ and $\sigma$ by minimizing $H_{\alpha, n}(\mu, \sigma)$ numerically. 
%We skip the rest of the mathematical details.

\subsubsection{Weibull distribution}\label{subsec:weibull}

Our final RM is the two-parameter Weibull distribution family, which has a distribution function 
$F_{(a,b)}(x) = 1 - \exp{\left[ - (b x)^a \right]}$ 
and density $f_{(a,b)}(x) = ab (bx)^{a-1} \exp{\left[ - (b x)^a \right]}$, for $x>0$, 
where $a$ and $b$ denote the shape and rate parameters, respectively.
Straightforward calculations again yield 
\begin{equation}\label{EQ:dpd5:weib}
   V_\alpha(a, b; x)  = \displaystyle \left\{\begin{array}{lll}
    \displaystyle \frac{a^\alpha b^\alpha \Gamma \left( 1 + \frac{(a-1) \alpha}{a} \right)}{(1 + \alpha)^{1 + \frac{(a-1) \alpha}{a}}} \\
   \displaystyle ~~~ - \left(1 + \frac{1}{\alpha}\right) a^\alpha b^\alpha (bx)^{\alpha(a-1)} \exp{\left[ - \alpha (b X_i)^a \right]},  & {\rm ~for} ~\alpha > 0,\\
	\displaystyle (b x)^a -\log(ab) - (a-1) \log(bx),  & {\rm ~for} ~\alpha = 0.  
\end{array}\right.
\end{equation}

Plugging in the sample observations $X_1, \ldots, X_n$ in (\ref{EQ:dpd5:weib}), 
we again calculate $H_{\alpha, n}(a, b) = \frac{1}{n} \sum_{i=1}^n V_\alpha(a, b; X_i)$ 
and then obtain the MDPDE of $a$ and $b$ by minimizing $H_{\alpha, n}(a, b)$ numerically. 
%We skip the rest of the mathematical details.

\subsection{Asymptotic relative efficiency} \label{subsec:are}

We study the performances of the proposed MDPDEs that can be expected through their theoretical properties. The first measure of the correctness of any estimator is its standard error or variance. 
{\color{black}It is difficult to obtain an analytic expression of the exact sampling distribution of MDPDEs, in general, and also in the case of MLE; however, certain asymptotic results are presented in \cite{basu1998robust}.} Let us assume that the model is correctly specified so that the true data generating distribution is $G=F_{\theta_0}$ for some $\theta_0\in \Theta$. Then, \cite{basu1998robust} prove that, under certain regularity conditions,  
$\widehat{\theta}_\alpha$ is a consistent estimator of $\theta_0$ and 
the asymptotic distribution of $\sqrt{n} (\widehat\theta_\alpha - \theta_0)$ is normal with mean zero and variance 
$J_\alpha(\theta_0)^{-1}K_\alpha(\theta_0) J_\alpha(\theta_0)^{-1}$, 
where
\begin{eqnarray}\label{EQ:J_K_xi}
&& J_\alpha(\theta) = \int u_\theta(x) u_\theta^T(x) f_\theta^{1+\alpha}(x) dx, \\
&& K_\alpha(\theta) = \int u_\theta(x) u_\theta^T(x) 
f_\theta^{1+2\alpha}(x) dx - \xi \xi^T, ~~ \xi = 
\int u_\theta(x) f_\theta^{1+\alpha}(x) dx.\nonumber
\end{eqnarray}
It is easy to verify that the asymptotic variance of the MDPDE is minimum when $\alpha = 0$. Thus, the asymptotic variance of the MDPDE is larger than that of the MLE, and considering an estimator with a smaller variance to be preferred in general, it is important to study the asymptotic relative efficiency (ARE), the ratio of the asymptotic variances of the MLE over that of the MDPDE assuming there is no outlier in the data. A value of ARE close to one indicates that the standard errors of the MLE and the MDPDE are comparable and hence, we can achieve robustness with only a little compromise in variance. By definition, the ARE of MDPDE at $\alpha=0$ is one for all models.

Among our RMs, if we consider the exponential model, the asymptotic variance of $\sqrt{n} (\widehat\lambda_\alpha - \lambda)$ 
can be computed to be $K / J^2$, where
\begin{eqnarray}\label{EQ:J_K_xi:exp}
&& J_\alpha(\lambda) = \frac{1 + \alpha^2}{(1 + \alpha)^3} \lambda^{\alpha - 2} \\
&& K_\alpha(\lambda) = \frac{1 + 4\alpha^2}{(1 + 2\alpha)^3} \lambda^{2\alpha - 2} - \xi^2, ~~ \xi = \frac{\alpha}{(1 + \alpha)^2} \lambda^{\alpha - 1} .\nonumber
\end{eqnarray}
Since the asymptotic variance of $\sqrt{n} (\widehat\lambda_{MLE} - \lambda)$ is $\lambda^{-2}$, 
the ARE of the MDPDE is then given by
\begin{equation}\label{EQ:dpd:are}
\textrm{ARE}(\widehat\lambda_\alpha) = \frac{\frac{1 + 4\alpha^2}{(1 + 2\alpha)^3} - \frac{\alpha^2}{(1 + \alpha)^4}}{\frac{(1 + \alpha^2)^2}{(1 + \alpha)^6}}.
\end{equation}
Note that it does not depend on the value of the parameter $\lambda$. 
For the other three RMs, however, the matrices $J_\alpha$ and $K_\alpha$ and hence the ARE of the MDPDEs cannot be computed explicitly and also depends on the underlying true parameter values; we compute them through numerical integrations. 
The ARE values obtained for the MDPDE at different $\alpha>0$ are presented in Table \ref{table_are} for all our RMs;
for the last three RMs, the results for some particular parameter values are presented. 

\begin{table}[ht]
\caption{Asymptotic relative efficiency of the MDPDEs at different $\alpha$ for the four RMs.}
\label{table_are}
\begin{tabular}{clccccccc}
\hline
&& \multicolumn{7}{c}{$\alpha$}\\
Distribution      & Par  & 0.1  & 0.2  & 0.3  & 0.4  & 0.5 & 0.7 & 1.0  \\
\hline
Exponential($\lambda$)       & $\lambda$ & 0.97 & 0.90 & 0.82 & 0.75 & 0.68 & 0.59 & 0.51 \\
\hline
Gamma(5, 0.05)    & $a$       & 0.98 & 0.94 & 0.88 & 0.82 & 0.77 & 0.68 & 0.58 \\
                  & $b$       & 0.98 & 0.93 & 0.86 & 0.80 & 0.74 & 0.64 & 0.55 \\
\hline
Gamma(10,0.05)    & $a$       & 0.98 & 0.93 & 0.87 & 0.81 & 0.75 & 0.66 & 0.56 \\
                  & $b$       & 0.98 & 0.93 & 0.86 & 0.79 & 0.73 & 0.64 & 0.54 \\
\hline                  
Weibull(2, 0.01)  & $a$       & 0.98 & 0.94 & 0.90 & 0.84 & 0.79 & 0.71 & 0.62 \\
                  & $b$       & 0.99 & 0.97 & 0.94 & 0.91 & 0.87 & 0.79 & 0.69 \\
\hline                   
Weibull(4, 0.01)  & $a$       & 0.99 & 0.94 & 0.88 & 0.82 & 0.78 & 0.69 & 0.59 \\
                  & $b$       & 0.99 & 0.97 & 0.93 & 0.90 & 0.86 & 0.78 & 0.67 \\
\hline                  
Lognormal(5, 0.2) & $\mu$     & 0.99 & 0.96 & 0.92 & 0.88 & 0.84 & 0.76 & 0.65 \\
                  & $\sigma$  & 0.98 & 0.92 & 0.85 & 0.79 & 0.73 & 0.63 & 0.54 \\
\hline                 
Lognormal(5, 0.4) & $\mu$     & 0.99 & 0.96 & 0.92 & 0.88 & 0.83 & 0.76 & 0.66 \\
                  & $\sigma$  & 0.98 & 0.92 & 0.85 & 0.78 & 0.72 & 0.63 & 0.54 \\
 \hline                 
\end{tabular}
\end{table}

{\color{black}From Table \ref{table_are}}, we observe from the table that the AREs of the MDPDEs decrease with increasing values of $\alpha$,
but the loss is not quite significant for smaller values of $\alpha>0$. Thus, the asymptotic variance of MDPDE under pure data (no outlier) assumption is comparable with that of the MLE at least for small values of $\alpha$ and, against this small price, we can achieve an extremely significant increase in the robustness under data contamination (when outliers are present), as illustrated in the next subsection. 

\subsection{Robustness: Influence function analysis}

We further illustrate the claimed robustness of the proposed MDPDE through the classical influence function analysis \citep{Hampeletc:1986}. For this purpose, we need to consider the functional approach with $T_\alpha(G)$ being the MDPDE functional at the true distribution $G$ as defined in (\ref{EQ:dpd2}) for tuning parameter $\alpha$.
Suppose $G_\epsilon =(1-\epsilon)G + \epsilon \wedge_y$ denotes the contaminated distribution where $\epsilon$ is the contamination proportion and $\wedge_y$ is the degenerate distribution at the contamination point (outlier) $y$.
Then, $\left[T_\alpha(G_\epsilon) - T_\alpha(G)\right]$ gives the (asymptotic) bias of the MDPDE due to contamination in data distribution. The influence function (IF) measures the standardized asymptotic bias of the estimator due to infinitesimal contamination and is defined as 
\begin{eqnarray}\label{EQ:if_def}
IF(y, T_\alpha, G) = \lim_{\epsilon\rightarrow 0}\frac{T_\alpha(G_\epsilon)-T_\alpha(G)}{\epsilon}.
\end{eqnarray}
Therefore, whenever the above IF is unbounded at the contamination point $y$, the bias of the underlying estimator can be extremely large (tending to infinity) even under infinitesimal contamination at a distant point;
this clearly indicates the non-robust nature of the corresponding estimator.
On the contrary, if the IF remains bounded in $y$, then the underlying estimator also remains within a bounded neighborhood of the true estimator even under contamination at far extreme $y$, and hence indicates the robustness of the estimator.

For our MDPDEs, the general theory developed in \cite{basu1998robust} can be followed to obtain its IF. 
When the model is correctly specified, i.e., $G=F_{\theta_0}$ for some $\theta_0\in\Theta$, 
the IF of the MDPDE functional $T_\alpha$ with tuning parameter $\alpha\geq 0$ is
\begin{eqnarray}\label{EQ:if_mdpde}
IF(y, T_\alpha, F_{\theta_0}) = J_\alpha(\theta_0)^{-1} \left[u_{\theta_0} (y) f_{\theta_0}^\alpha(y) - \int u_{\theta_0}(x) f_{\theta_0}^{1+\alpha}(x) dx\right],
\end{eqnarray}
where $J_\alpha$ is as defined in (\ref{EQ:J_K_xi}). Due to the exponential nature of the densities of our RMs, it can be shown that the corresponding score functions $u_\theta$ are polynomial; 
hence the IF of the MDPDEs for our RMs are bounded for all $\alpha>0$ but unbounded at $\alpha=0$ (corresponds to the MLE).

\begin{figure}[h]
	\centering
	\adjincludegraphics[width = 0.45\linewidth, trim = {{.0\width} {.0\width} {.0\width} {.0\width}}, clip]{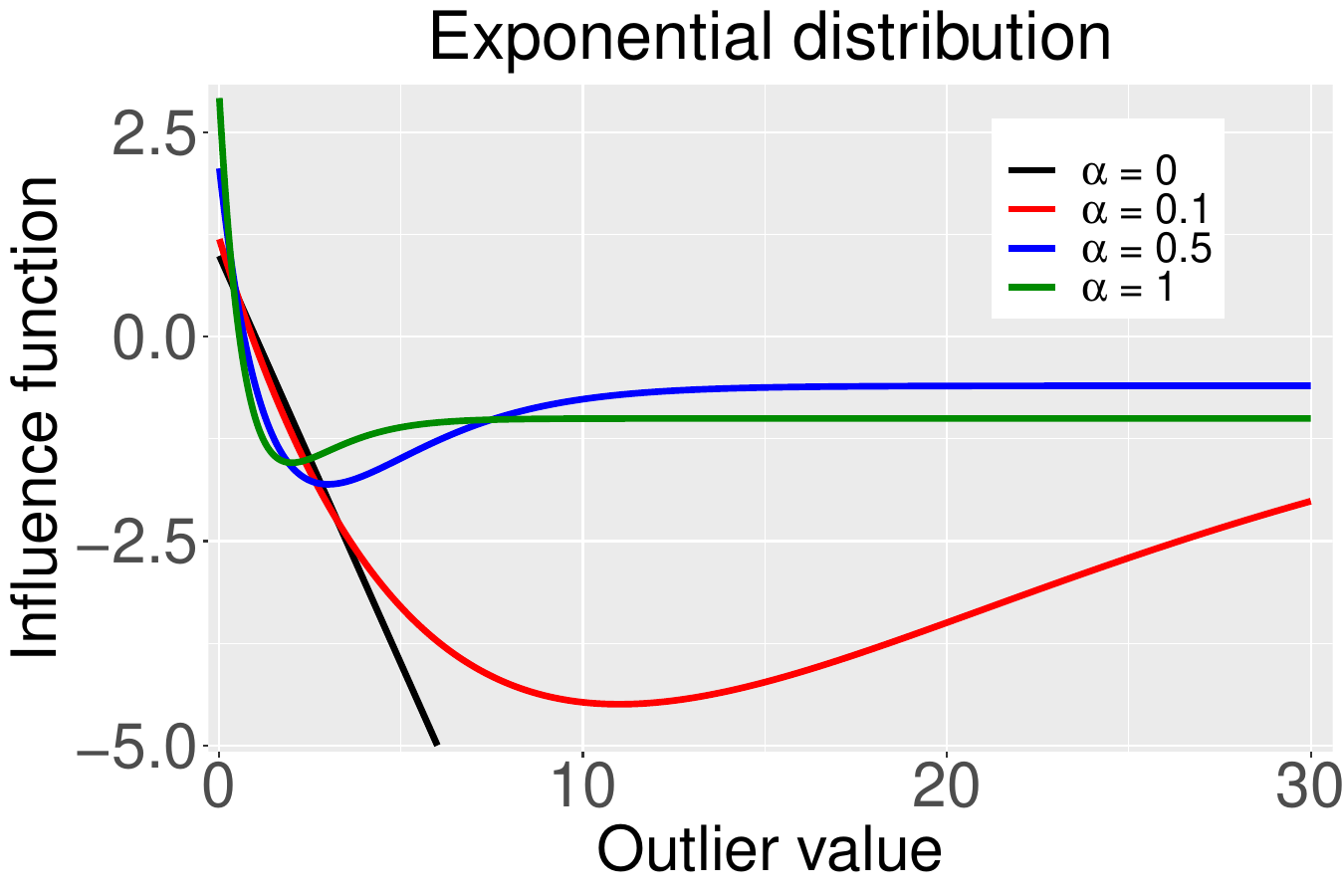}
\adjincludegraphics[width = 0.45\linewidth, trim = {{.0\width} {.0\width} {.0\width} {.0\width}}, clip]{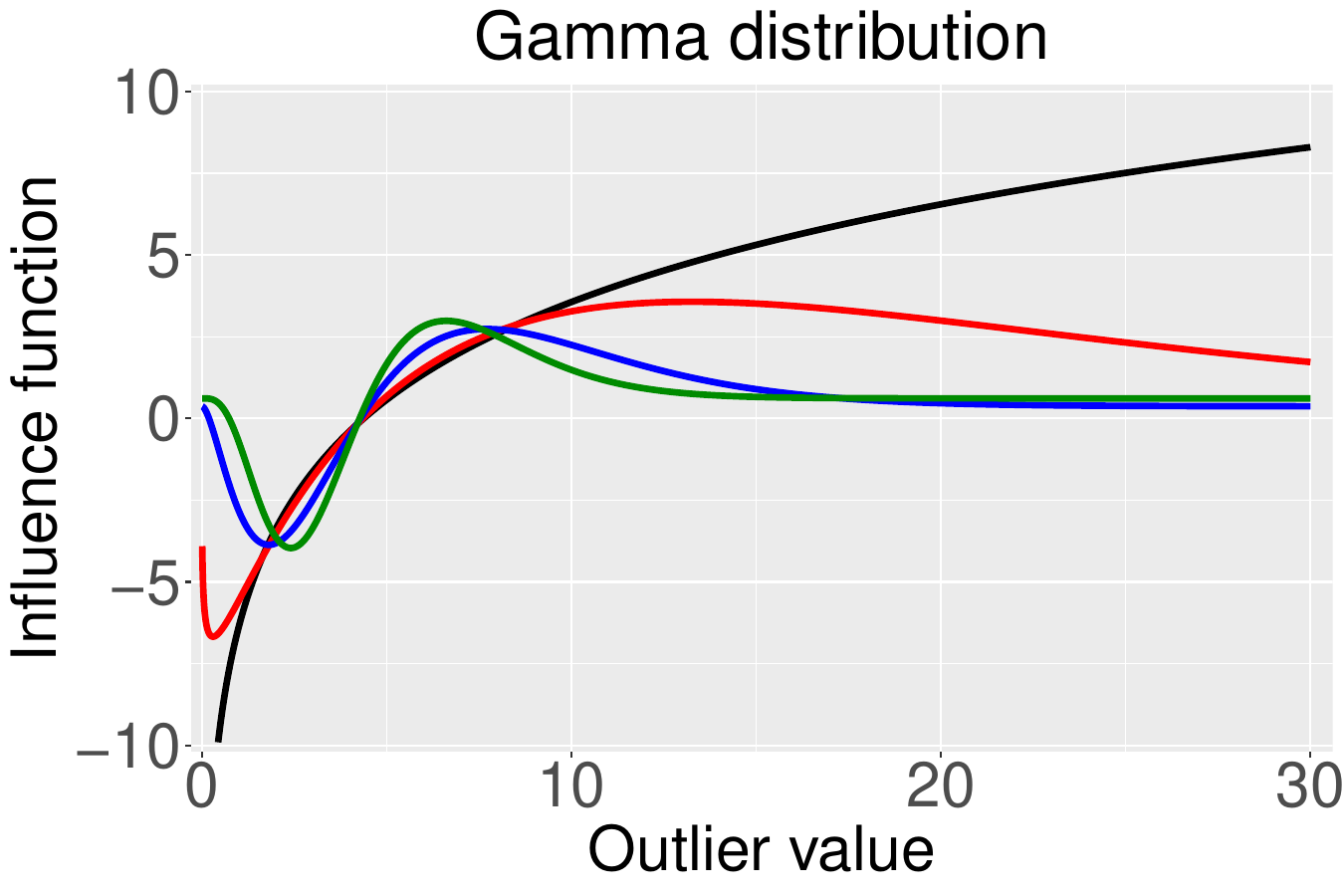}
\adjincludegraphics[width = 0.45\linewidth, trim = {{.0\width} {.0\width} {.0\width} {.0\width}}, clip]{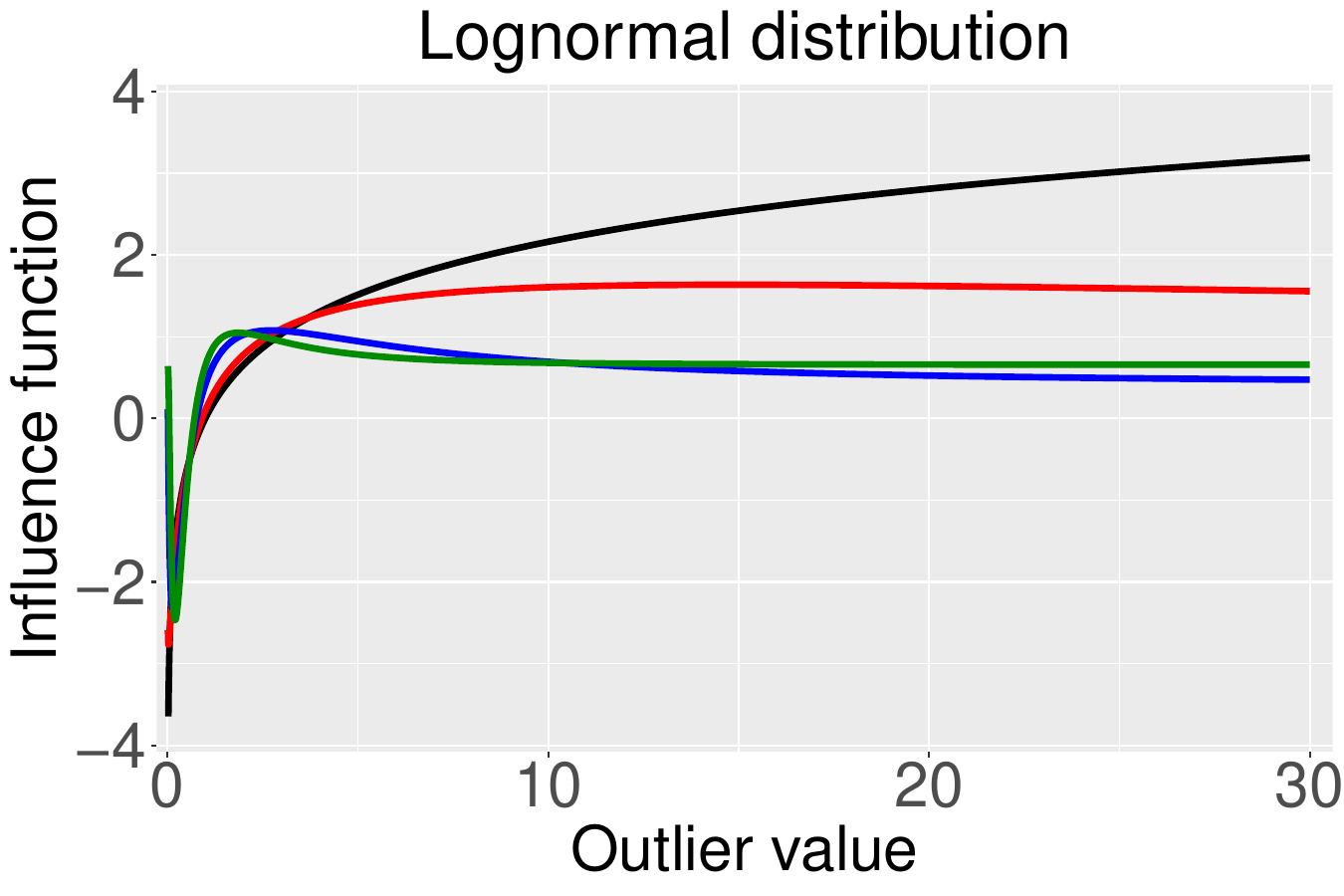}
\adjincludegraphics[width = 0.45\linewidth, trim = {{.0\width} {.0\width} {.0\width} {.0\width}}, clip]{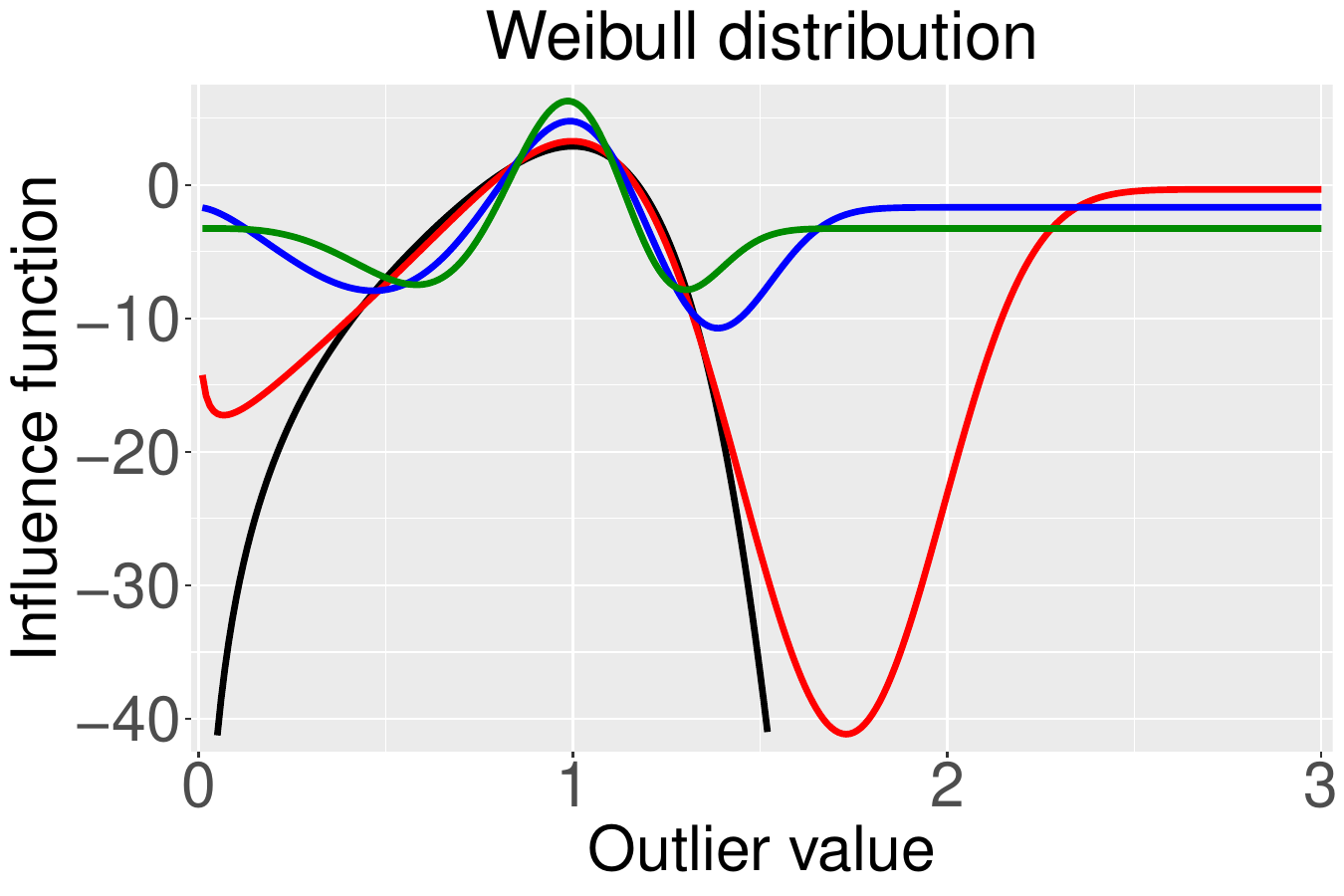}
\caption{Influence Functions of the MDPDEs for different choices of the outlier value $y$ in \eqref{EQ:if_def} for different RMs. All the sub-figures share the same legend as in the top-left panel. The top-left panel describes the IF for $\lambda$ of exponential($\lambda$) at exponential(1) distribution. The top-right panel describes the IF for $a$ of gamma($a,1$) at gamma($5,1$) distribution. The bottom-left panel describes the IF for $\mu$ of lognormal($\mu, 1$) at lognormal($0, 1$) distribution. The bottom-right panel describes the IF for $a$ of Weibull($a,1$) at Weibull($5,1$) distribution.}
	\label{fig_if}
\end{figure}

To visualize it more clearly, in Figure \ref{fig_if}, we present the IFs of the MDPDEs, at different $\alpha$, over the contamination point $y$ for the four RMs with some particular values of model parameters $\theta_0$. Specifically, in the top-left panel, we describe the IF for the rate parameter $\lambda$ of the exponential distribution with true rate parameter $\lambda_0 = 1$. The top-right panel describes the IF for the shape parameter $a$ of the gamma distribution with true shape parameter $a_0 = 5$ when the rate parameter $b=1$ is fixed. The bottom-left panel describes the IF for the parameter $\mu$ of the lognormal distribution with true value $\mu_0 = 0$ when the parameter $\sigma_0=1$ is fixed. The bottom-right panel describes the IF for the shape parameter $a$ of the Weibull distribution with true shape parameter $a_0 = 5$ when the rate parameter $b=1$ is fixed. The boundedness of the IFs at $\alpha>0$ and their unboundedness at $\alpha=0$ are observed from the figures. This indicates the claimed robustness of the MDPDEs at $\alpha>0$, and the non-robust nature of the MLE (at $\alpha=0$).

\subsection{On the choice of optimum tuning parameter $\alpha$}

We have seen that the tuning parameter $\alpha$ provides a trade-off between the efficiency and robustness of the corresponding MDPDEs; choosing a larger $\alpha$ provides higher robustness while the asymptotic variance (and hence the standard error) of the estimator increases. So one needs to choose $\alpha$ appropriately depending on the amount of contamination in the data; larger $\alpha$ for greater contamination proportions and vice versa. In practice, however, the contamination proportion in the data is unknown (can be guessed at maximum) and hence a data-driven algorithm for the selection of an `optimum' tuning parameter $\alpha$ is necessary for practical applications of the MDPDE including the present rainfall modeling.

% and \cite{seo2017extreme} use it for extreme rainfall modeling. 

A leave-one-out cross-validation approach is proposed by \cite{fujisawa2006robust} for selecting `optimum' $\alpha$ in MDPDE. Suppose we observe the IID samples $X_1, \ldots, X_n$ and assume that they follow a distribution function $F_{\bm{\theta}}$ parameterized by $\bm{\theta}$. For a specific choice of $\alpha$, for each $i \in \{ 1, \ldots, n\}$, we estimate $\bm{\theta}$ based on all samples except $X_i$ and let it be denoted by $\widehat{\bm{\theta}}_\alpha^{(-i)}$. Further, the authors calculate the empirical Cramer-von Mises (CVM) distance $n^{-1} \sum_{i=1}^{n} \lbrace (i - 0.5)/n - F_{\widehat{\bm{\theta}}_\alpha^{(-i)}}(X_{(i)}) \rbrace^2$ and minimize it over $\alpha$ to obtain the `optimum' tuning parameter. Here $X_{(1)}, \ldots, X_{(n)}$ denote the order statistics of the sample observations. However, we replace the CVM distance with the more robust Wasserstein distance (WD) while keeping the leave-one-out cross-validation approach of \cite{fujisawa2006robust}. Here, we choose the `optimum' tuning parameter by minimizing the empirical WD distance over $\alpha$ as %following the suggestion made by an anonymous reviewer,
\begin{equation}\label{EQ:AD}
\alpha^\ast = \arg \min_{\alpha} \frac{1}{n} \sum_{i=1}^{n} \left| \frac{i - 0.5}{n} - F_{\widehat{\bm{\theta}}_\alpha^{(-i)}}(X_{(i)}) \right|.
\end{equation}

\subsection{Robust Model Selection}\label{model_selection}

The final step in rainfall modeling is to choose an appropriate parametric model from a set of candidate RMs. The usual approach of the Akaike information criterion (AIC) is based on the MLE and hence non-robust against outliers. In consistency with the robust MDPDE, we use an associated robust model selection criterion, namely the robust information criterion (RIC).

Although first discussed in the technical report associated with \cite{basu1998robust}, RIC has been widely explored in \cite{Mattheou2009}.  
For a particular model $M$, if $\widehat{\theta}_{\alpha,M}$ denote the MDPDE at some prefixed tuning parameter $\alpha \geq 0$,
then the corresponding value of RIC is computed as 
\begin{equation}\label{EQ:RIC}
RIC_{\alpha, M} = H_{\alpha, n}\left(\widehat{\theta}_{\alpha,M}\right) 
+ (1 + \alpha)^{-1} n^{-1} Tr\left[J_\alpha(\widehat{\theta}_{\alpha,M})^{-1} K_\alpha(\widehat{\theta}_{\alpha,M})\right],
\end{equation}
where $H_{\alpha, n}$ is the MDPDE objective function given in \eqref{EQ:dpd4}, the matrices $J_\alpha$ and $K_\alpha$ are as defined in (\ref{EQ:J_K_xi}) and $Tr(\cdot)$ denotes the trace of a matrix. For fixed $\alpha\geq 0$, the RIC values are compared across different models, and the model having the lowest RIC is chosen as the ``best" among the candidate RMs. It is important to note that, at $\alpha = 0$, MDPDE and MLE coincide, and hence the model with minimum RIC is the same as the model with minimum AIC. 

{\color{black}One downside of RIC is that it depends on the tuning parameter $\alpha$. As a result, for each model $M$, we choose $\alpha$ first by minimizing the WD as described in the previous subsection and then apply RIC with the chosen `optimum' $\alpha$, say $\alpha^\ast_M$ to select the final rainfall model in each case robustly. Therefore, for rainfall model selection, we can select a model $M^*$ satisfying 
$$M^* = \arg\min_{M} RIC_{\alpha^\ast_M, M}.$$}

{\color{black}The model selection and tuning parameter selection have been treated differently and never together in the literature as of our knowledge.
Further theoretical or numerical studies judging the goodness-of-fit of a model $M^\ast$ are currently unavailable in the literature and such a comparison is out of the scope of this research. However, which rainfall model to use for modeling the detrended rainfall data from a specific subdivision-month combination is a natural and crucial question. In this context, intuitively, $RIC_{\alpha^\ast_M, M}$ seems the most reasonable representative of RIC for model $M$ assuming \eqref{EQ:AD}, and further, it again seems natural to compare such representatives of RIC values to select a final model. Certain visual inspections should be done to judge the accuracy of selecting a correct model along with comparing based on RIC; e.g., if the data exhibits a mode at a nonzero value, it is unlikely that an exponential distribution would reasonably explain the underlying distribution. If $\alpha_M^\ast$ are approximately the same for all $M$, a model selection based on RIC is theoretically reliable.}

\subsection{$L$-moments estimation}\label{l_moment}

The $L$-moments estimation (LME, henceforth), proposed by \cite{hosking1990moments}, is a widely used tool for robust inference in hydrology. In the method of moments estimation, we obtain the parameter estimates by equating the raw/central population moments and the sample moments; however, in LME, we equate the population moments and the sample moments of some linear combinations of order statistics and solve the corresponding equations to obtain the estimates. These linear combinations of order statistics are called $L$-statistics. The LMEs have some theoretical advantages over conventional moments for being more robust to outliers. In small sample scenarios, LMEs usually provide less bias compared to MLEs in the presence of outliers.

Here, the family of rainfall models we consider, i.e., RM, includes a one-parameter distribution (exponential) and three two-parameter distributions (gamma, Weibull, lognormal). Thus, for the first one, we equate only the first theoretical and sample $L$-moments. In other cases, we equate the first two theoretical and sample $L$-moments. If $X_{(1)}, \ldots, X_{(n)}$ denote the ordered sample, then we obtain the first and second sample $L$-moments as
% using the same notation as \cite{hosking1990moments}, 
\begin{equation} \label{lme_sample}
    l_1 = \frac{1}{n} \sum_{i=1}^n X_{(i)}, ~~~~ l_2 = \frac{2}{n(n-1)} \sum_{i=1}^n (i-1) X_{(i)} - l_1,
\end{equation}
and the first two population $L$-moments are
\begin{equation} \label{lme_population}
    L_1 = \int_0^\infty x f_{\bm{\theta}}(x) dx, ~~~~ L_2 = 2 \int_0^\infty x F_{\bm{\theta}}(x)  f_{\bm{\theta}}(x) dx - L_1,
\end{equation}
where $F_{\bm{\theta}}(\cdot)$ and $f_{\bm{\theta}}(\cdot)$ are generic notations for the distribution functions and the density functions of the members of RM.

For the exponential distribution in Section \ref{subsec:exponential}, $L_1 = \lambda^{-1}$ and $l_1 = \Bar{X}$, the sample mean, follows from \eqref{lme_sample} and \eqref{lme_population}. Thus, the LME of $\lambda$ is given by $\widehat{\lambda}_{\textrm{LME}} = \Bar{X}^{-1}$, which is identical to the MLE in this case. For the other members of RM, LMEs and MLEs are not identical. In these cases, we calculate the integral in $L_2$ in \eqref{lme_population} numerically.

%\vspace{-2mm}
\section{Results}
\label{results}

To illustrate the advantages of the proposed MDPDE in rainfall modeling, we first present the histograms of {\color{black} detrended} rainfall data along with fitted models by MDPDE (and MLE and LME) for four example cases of subdivision-month combinations under each member of RM. For this purpose, we choose four values of $\alpha = 0, 0.1, 0.5, 1$, and discuss the corresponding MDPDE estimates, along with the standard error (SE) and the Wasserstein distances (WD) for each case. {\color{black} Along with these arbitrary choices of $\alpha$, we also discuss the results based on `optimal' $\alpha$, denoted by $\alpha^\ast$ in \eqref{EQ:AD}}. Because MDPDE for $\alpha=0$ coincides with the case of MLE, we do not discuss MLE separately. For LME, we discuss the estimators, their SEs, and the corresponding WDs as well. {\color{black}Due to certain ambiguities with a simultaneous selection of the underlying probability model and the tuning parameter as mentioned in Section \ref{model_selection}, we choose certain representative examples based on visual inspections of the histograms and postpone the discussion on RIC-based model selection to Section \ref{subsec:final_selection}}.

The overall results from the comprehensive study are presented afterward along with the final predicted models and the estimated median rainfall amounts for all month-subdivision combinations.

\subsection{{\color{black}MDPDE with exponential distribution}}

%  (Gujarat region; May), 

\begin{figure}[h]
    \centering
    \adjincludegraphics[width = 0.45\linewidth, trim = {{.0\width} {.0\width} {.0\width} {.0\width}}, clip]{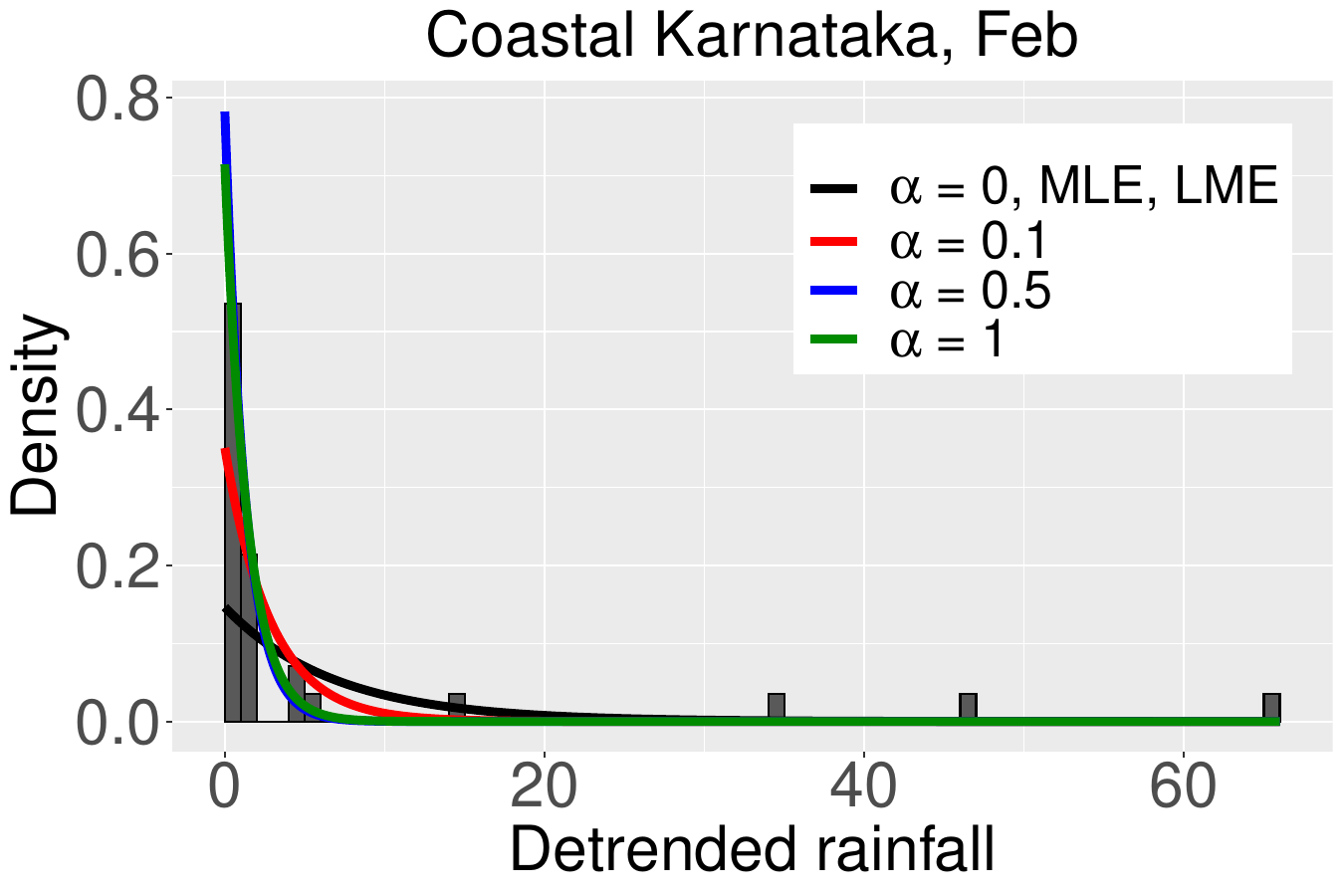}
\adjincludegraphics[width = 0.45\linewidth, trim = {{.0\width} {.0\width} {.0\width} {.0\width}}, clip]{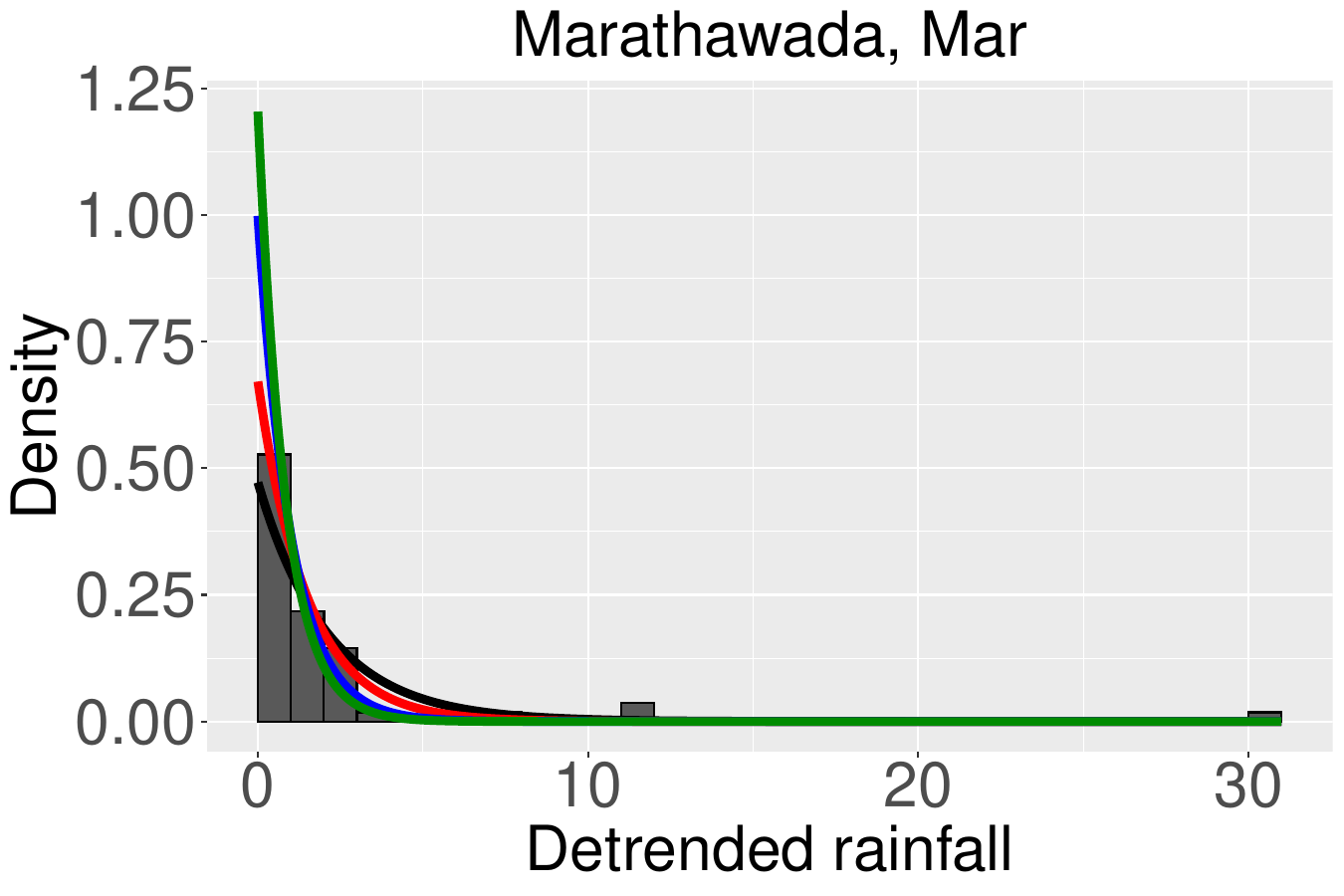} \\
\adjincludegraphics[width = 0.45\linewidth, trim = {{.0\width} {.0\width} {.0\width} {.0\width}}, clip]{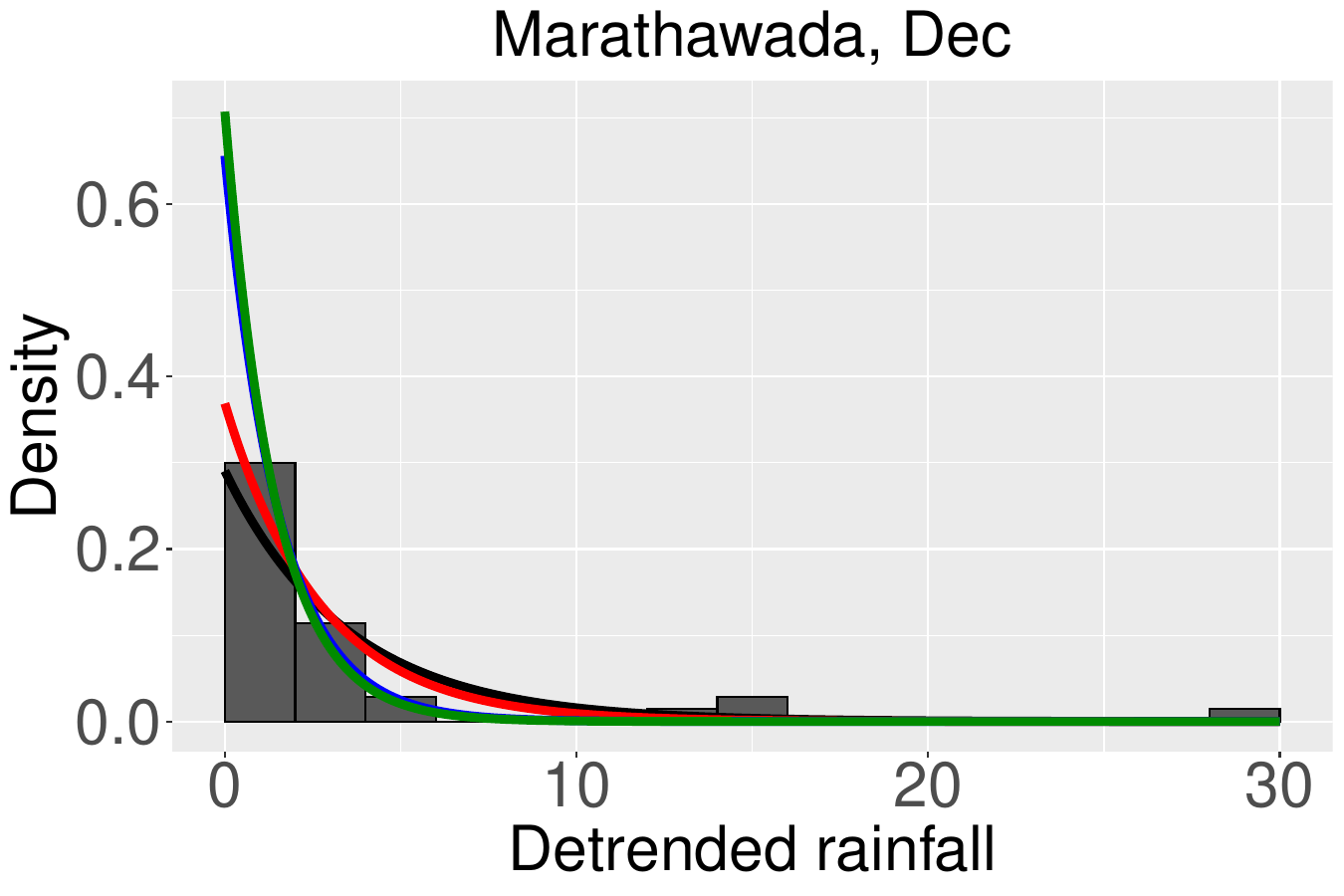}
\adjincludegraphics[width = 0.45\linewidth, trim = {{.0\width} {.0\width} {.0\width} {.0\width}}, clip]{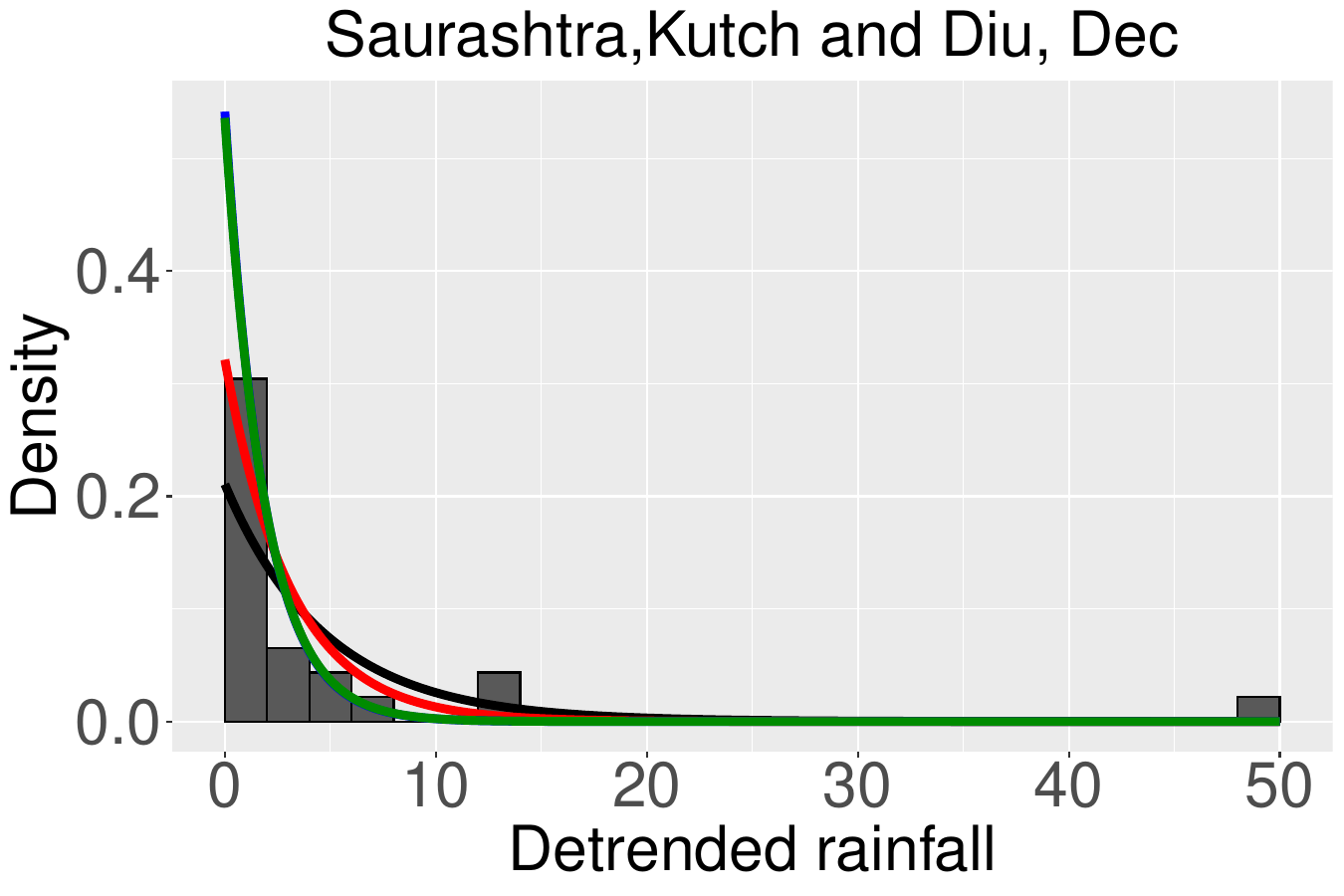}
    \caption{Histograms along with fitted exponential densities {\color{black} to detrended rainfall data} with parameters estimated using the MDPDE at $\alpha = 0, 0.1, 0.5, 1$. The MLE and LME results coincide with the MDPDE case at $\alpha=0$. All the sub-figures share the same legend as in the top-left panel.}
    \label{fig_exponential_bestfits}
\end{figure}

We choose four subdivision-month combinations where exponential distribution provides more reliable fits to {\color{black} detrended rainfall data} compared to other members of RM. The density function of the exponential distribution is monotonically decreasing which resembles the empirical histograms for these cases; also these data include a high percentage of outliers. These chosen cases are -- (Coastal Karnataka; February), {\color{black}(Marathwada; March)}, (Marathwada; December), and (Saurashtra, Kutch, and Diu; December), having outlier proportions of {\color{black}10.7\%, 5.5\%, 2.9\%, and 26.1\%}, respectively. The fitted exponential distributions based on different estimation approaches and tuning parameter settings are provided in Figure \ref{fig_exponential_bestfits}. For exponential distribution, LME coincides with MLE. %and MDPDE for $\alpha=0$
The corresponding MDPDEs, their SEs, and the WDs are provided in Table \ref{table_exponential_bestfits}.

\begin{table}[h]
\caption{{\color{black}Parameter estimates, corresponding standard errors, and empirical Wasserstein distances (WD) for fitting exponential distributions to detrended rainfall data at selective subdivision-month pairs, using MDPDE, MLE, and LME. The values of $\alpha^\ast$ for the four subdivision-month combinations are 0.9913, 0.1548, 0.1841, and 0.2869, respectively.}}
\label{table_exponential_bestfits}
\centering
{\color{black}
\begin{tabular}{ccccccc}
\hline
\begin{tabular}[c]{@{}c@{}}Region, \\ Month\end{tabular} & Results      & $\alpha = 0$/MLE/LME & $\alpha = 0.1$ & $\alpha = 0.5$ & $\alpha = 1$ & $\alpha = \alpha^\ast$ \\
\hline
Coastal  & $\hat{\lambda}$ & 0.1466 & 0.3507 & 0.7819 & 0.7143 & 0.7154 \\ 
Karnataka,   & $\textrm{SE}(\hat{\lambda})$ & 0.1164 & 0.2781 & 0.1774 & 0.1497 & 0.1500 \\ 
February    & WD  & 0.2454 & 0.1422 & 0.0596 & 0.0533 & 0.0533 \\
\hline  
Maratwada,  & $\hat{\lambda}$   & 0.4720 & 0.6711 & 0.9982 & 1.2040 & 0.7591 \\ 
March   & $\textrm{SE}(\hat{\lambda})$ & 0.1434 & 0.1600 & 0.3147 & 0.5744 & 0.1742 \\ 
    & WD & 0.1313 & 0.0657 & 0.0878 & 0.1135 & 0.0559 \\ 
\hline
Maratwada, & $\hat{\lambda}$ & 0.2905 & 0.3688 & 0.6560 & 0.7072 & 0.4668 \\ 
December   & $\textrm{SE}(\hat{\lambda})$ & 0.1106 & 0.1377 & 0.2303 & 0.2597 & 0.1739 \\ 
  & WD  & 0.1352 & 0.0921 & 0.0700 & 0.0734 & 0.0666 \\ 
\hline
Saurashtra,  & $\hat{\lambda}$ & 0.2106 & 0.3213 & 0.5416 & 0.5360 & 0.4804 \\ 
Kutch \& Diu,  & $\textrm{SE}(\hat{\lambda})$ & 0.1364 & 0.1616 & 0.1753 & 0.1538 & 0.1759 \\ 
December  & WD & 0.1602 & 0.1015 & 0.0577 & 0.0559 & 0.0533 \\ 
   \hline
\end{tabular}
}
\end{table}

% In an asymptotic sense, the standard errors of the MDPDEs are proportional to the corresponding rate parameters (follows from (\ref{EQ:J_K_xi:exp})), and hence they also increase with $\alpha$. Here, we observe a similar pattern for most of the cases. 

{\color{black}From Figure \ref{fig_exponential_bestfits}, we observe that the fitted density based on MLE or LME (black line; $\alpha = 0$) has a thicker right tail, and with increasing $\alpha$, the tails of the fitted densities become thinner. 
Hence, the fitted densities underestimate the probabilities of smaller values and overestimate the probabilities of incorrectly large values (outliers) for $\alpha = 0$ (the MLE-based and LME-based inference). On the other hand, for $\alpha=1$, 
the fitted densities appear to overestimate the probability of smaller values and underestimate the probability of moderate through large values; this is due to high penalization (assigning less probability) of the moderate values along with the outliers. There is a clear trend for over-and-underestimation as we move from $\alpha=0$ to $\alpha=1$. This fact can also be observed from the third through sixth columns of Table \ref{table_exponential_bestfits}, where the MDPDE estimates increase with increasing $\alpha$ in most cases; note that the tail of an exponential distribution becomes thinner with increasing value of its rate parameter. The standard errors of the MDPDEs also increase with $\alpha$ in most cases. Further, the WDs decrease with changing $\alpha = 0$ to $\alpha = 0.1$ for all the cases, explaining the reduction in bias with increasing $\alpha$. For the first subdivision-month combination, $\alpha^\ast$ is close to one and the corresponding WD is also close to that based on $\alpha=1$. For the other three combinations, $\alpha^\ast$ values are moderate and the corresponding $\hat{\lambda}$ and $\textrm{SE}(\hat{\lambda})$ values are also in between their values based on the extreme choices $\alpha=0$ and $\alpha=1$ in most cases indicating a balance in the bias-variance trade-off.}

% ; from the topleft panel of Figure \ref{fig_exponential_bestfits}, several 
% Based on the minimization of the WDs, the optimal $\alpha$ values for the considered subdivision-month pairs turn out to be 0.2558, 0.9980, 0.6281, and 0.3022, respectively, with the corresponding WDs being 0.0879, 0.0680, 0.0708, and 0.0549, respectively. The MDPDEs for these final `best' fitted models are 0.2066, 1.8792, 0.1846, and 0.9028, respectively, with the corresponding standard errors being 0.1004, 0.4309, 0.0672, and 0.3259, respectively.

% Therefore, an optimal $\alpha$ between 0.1 and 1 would provide the best fit for the data by controlling the penalization based on the number of outliers in each case. 

%\vspace{-2mm}
\subsection{{\color{black} MDPDE with gamma distribution}}

\begin{figure}[h]
    \centering
   \adjincludegraphics[width = 0.45\linewidth, trim = {{.0\width} {.0\width} {.0\width} {.0\width}}, clip]{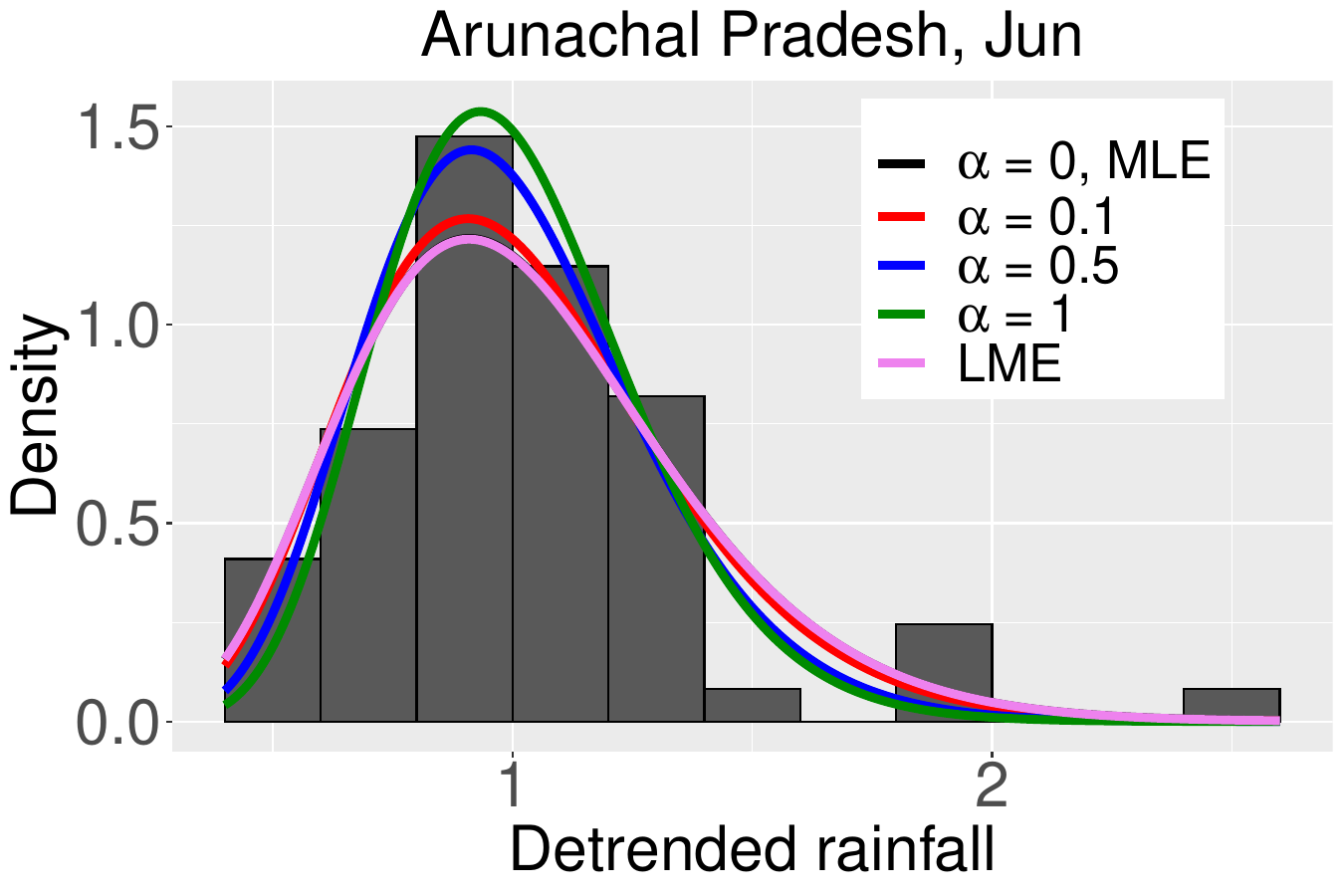}
    \adjincludegraphics[width = 0.45\linewidth, trim = {{.0\width} {.0\width} {.0\width} {.0\width}}, clip]{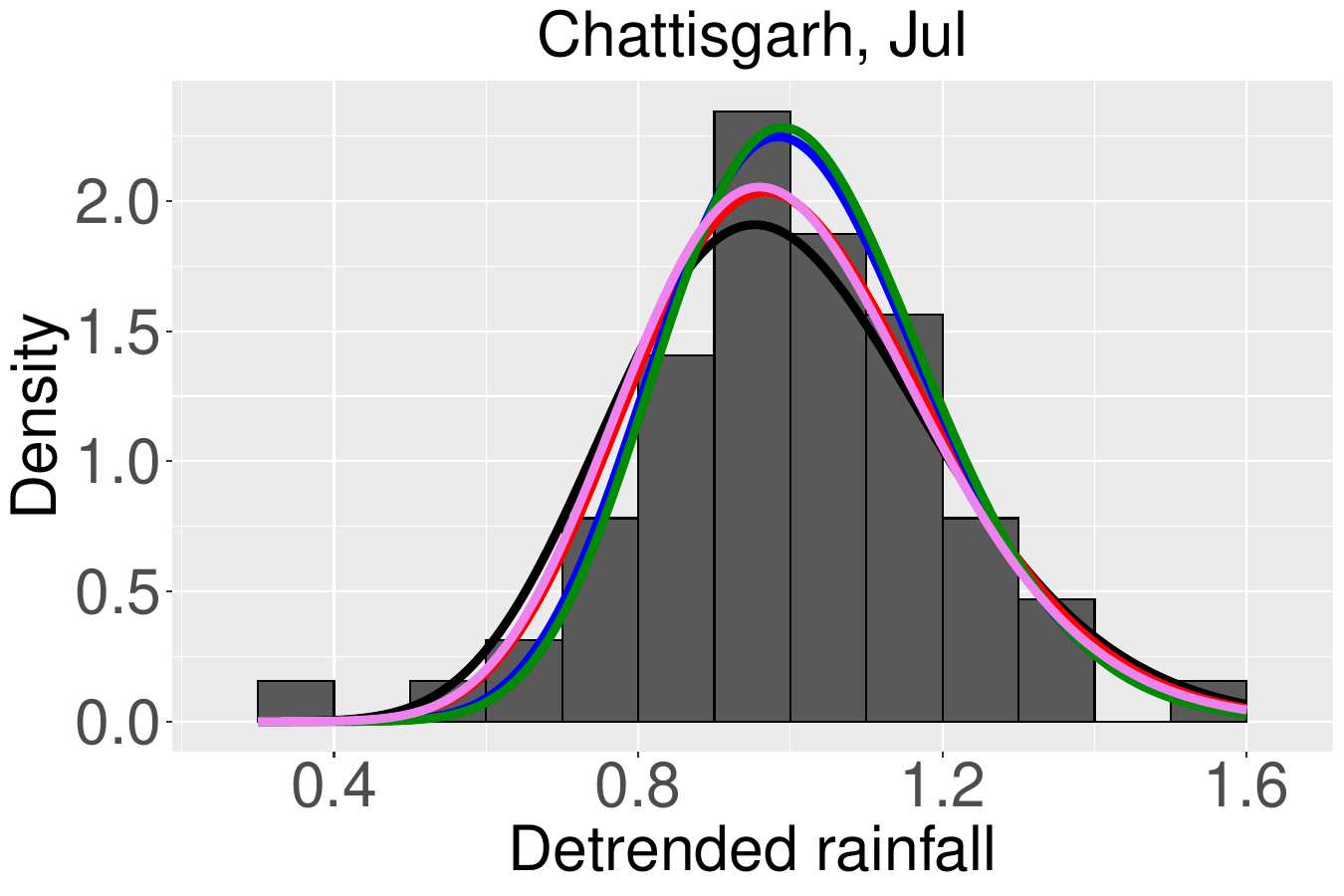} \\
    \adjincludegraphics[width = 0.45\linewidth, trim = {{.0\width} {.0\width} {.0\width} {.0\width}}, clip]{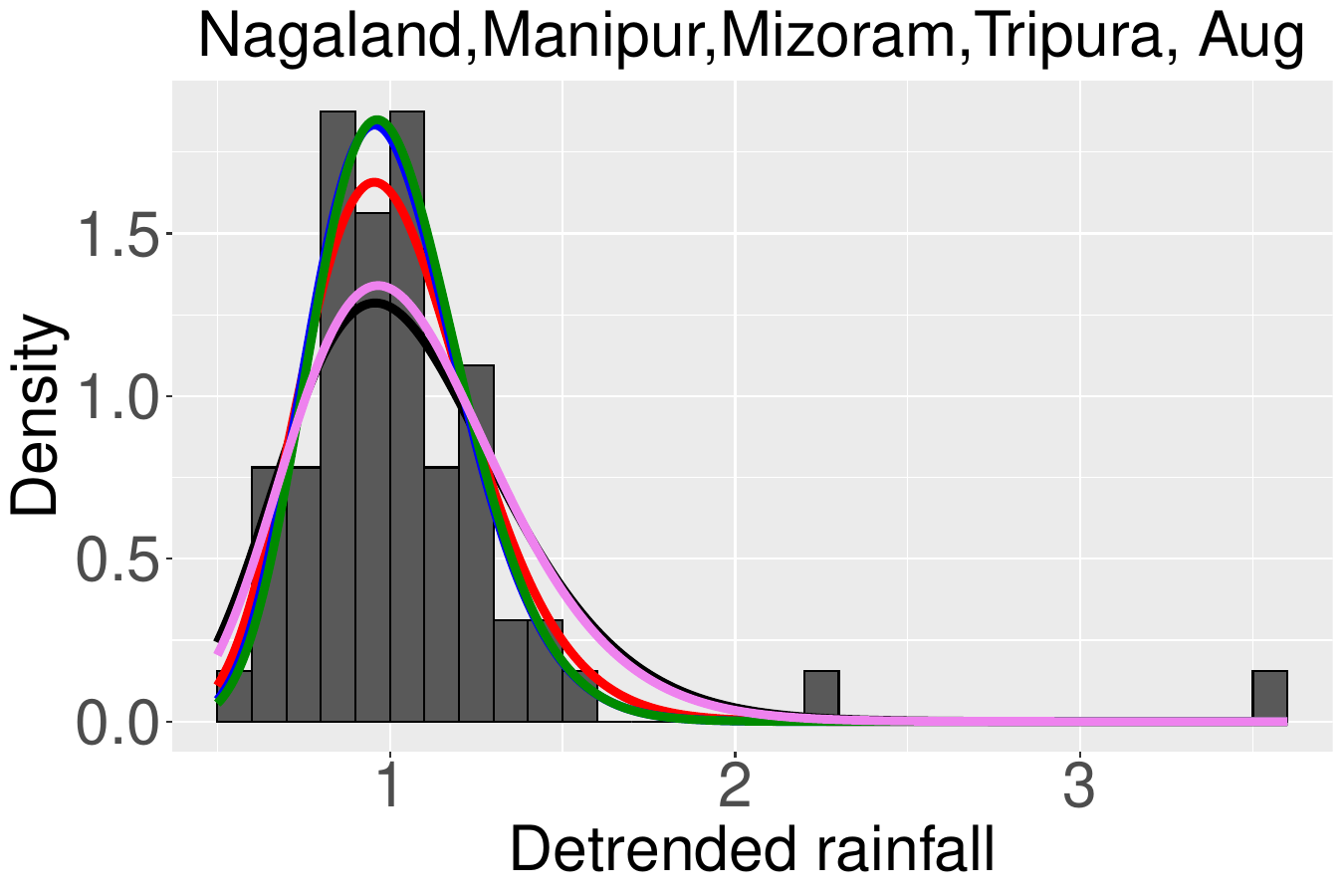}
    \adjincludegraphics[width = 0.45\linewidth, trim = {{.0\width} {.0\width} {.0\width} {.0\width}}, clip]{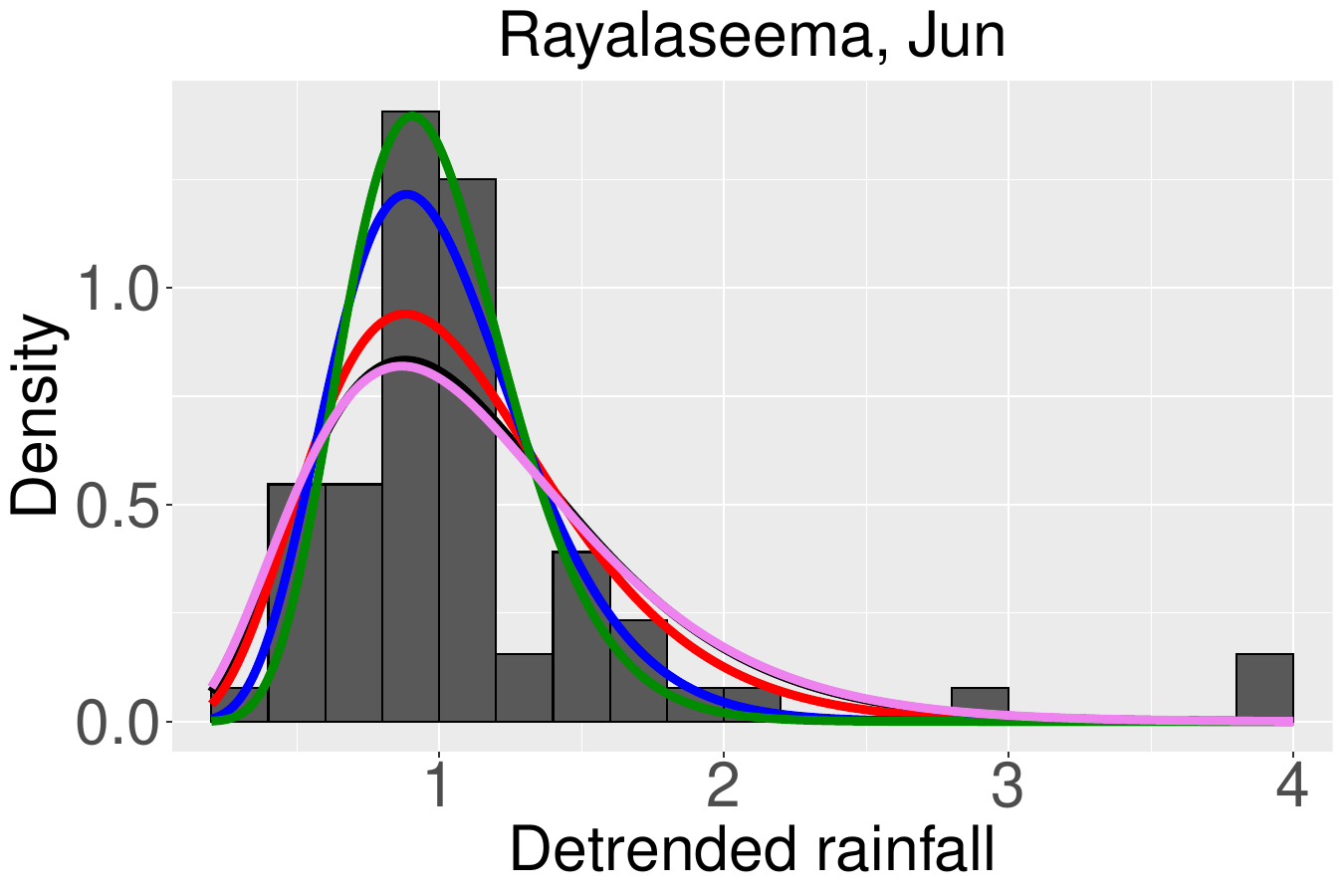}
    \caption{Histograms along with fitted gamma densities {\color{black} to detrended rainfall data} with parameters estimated using MDPDE at $\alpha = 0, 0.1, 0.5, 1$, and LME. The results for MLE coincide with the case of MDPDE at $\alpha=0$. All the sub-figures share the same legend as in the top-left panel}
    \label{fig_gamma_bestfits}
\end{figure}

Here we choose four subdivision-month combinations where gamma distribution provides reliable fits and the data include some outliers. The selected cases are -- (Arunachal Pradesh; June), (Chattisgarh; July), (Nagaland, Manipur, Mizoram, and Tripura; August), and (Rayalseema; June); the proportions of outliers in these cases are {\color{black}8.2\%, 3.1\%, 3.1\%, and 7.8\%}, respectively. The fitted gamma densities based on MDPDE and LME are provided in Figure \ref{fig_gamma_bestfits}, while the corresponding MDPDEs, their SEs, and WDs are provided in Table \ref{table_gamma_bestfits}.

\begin{table}[ht]
\caption{{\color{black} Parameter estimates, corresponding standard errors, and empirical Wasserstein distances (WD) for fitting gamma distributions to detrended rainfall data at selective subdivision-month pairs, using MDPDE, MLE, and LME. The values of $\alpha^\ast$ for the four subdivision-month combinations are 0.2562, 0.2305, 0.1277, and 0.1798, respectively.}}
\label{table_gamma_bestfits}
\centering
{\color{black}
\begin{tabular}{cccccccc}
\hline
\begin{tabular}[c]{@{}c@{}}Region, \\ Month\end{tabular} & Results  & $\alpha = 0$/MLE   & $\alpha = 0.1$    & $\alpha = 0.5$    & $\alpha = 1$ & LME & $\alpha = \alpha^\ast$ \\
\hline
Arunachal  & $\hat{a}$ & 8.8730 & 9.4768 & 12.0728 & 14.1102 & 8.8428 & 11.1604 \\ 
Pradesh,  & $\hat{b}$ & 8.6573 & 9.3429 & 12.1128 & 14.0458 & 8.6277 & 11.1760 \\ 
June  & $\textrm{SE}(\hat{a})$ & 2.3220 & 2.5605 & 5.0220 & 6.2096 & 2.6283 & 4.0834 \\ 
 & $\textrm{SE}(\hat{b})$ & 2.4824 & 2.6866 & 5.0309 & 6.1629 & 2.7811 & 4.1179 \\ 
 & WD  & 0.0299 & 0.0252 & 0.0199 & 0.0237 & 0.0299 & 0.0196 \\
  \hline
Chattisgarh,  & $\hat{a}$ & 21.9655 & 25.4506 & 31.8890 & 33.2228 & 25.5614 & 29.4809 \\ 
July  & $\hat{b}$ & 21.9977 & 25.2932 & 31.3875 & 32.5661 & 25.5988 & 29.1273 \\ 
  & $\textrm{SE}(\hat{a})$ & 7.1828 & 7.1361 & 9.1744 & 10.7913 & 6.3370 & 7.8868 \\ 
    & $\textrm{SE}(\hat{b})$ & 6.9399 & 6.9185 & 8.9376 & 10.5141 & 6.1210 & 7.6633 \\ 
  & WD & 0.0322 & 0.0204 & 0.0170 & 0.0227 & 0.0244 & 0.0142 \\
  \hline
Nagaland,  & $\hat{a}$ & 10.6993 & 16.9058 & 20.4476 & 21.1001 & 11.6895 & 21.0836 \\ 
Manipur,  & $\hat{b}$ & 10.1312 & 16.6597 & 20.3568 & 20.8768 & 11.0689 & 20.8623 \\ 
Mizoram,  & $\textrm{SE}(\hat{a})$ & 5.3595 & 4.9638 & 5.8830 & 8.4364 & 5.1984 & 8.4029 \\ 
and Tripura,  & $\textrm{SE}(\hat{b})$ & 5.5746 & 5.0951 & 5.8219 & 8.2954 & 5.4092 & 8.2632 \\ 
August  & WD & 0.0528 & 0.0213 & 0.0210 & 0.0193 & 0.0480 & 0.0193 \\
  \hline
Rayalseema,  & $\hat{a}$ & 4.5385 & 5.4669 & 8.4516 & 11.1996 & 4.3471 & 6.3613 \\ 
June  & $\hat{b}$ & 4.0211 & 5.0709 & 8.4104 & 11.2657 & 3.8517 & 6.0807 \\ 
  & $\textrm{SE}(\hat{a})$ & 1.3355 & 1.6257 & 4.8624 & 8.2844 & 1.5027 & 2.0315 \\ 
  & $\textrm{SE}(\hat{b})$ & 1.4487 & 1.7573 & 5.0859 & 8.5417 & 1.5980 & 2.1847 \\ 
     & WD  & 0.0570 & 0.0414 & 0.0416 & 0.0415 & 0.0581 & 0.0382 \\ 
   \hline
\end{tabular}
}
\end{table}

%  but still has a similar pattern of underestimation and overestimation as for the MLE
{\color{black}From the histograms in Figure \ref{fig_gamma_bestfits}, we notice that the outliers are on the right tail except for the second case where the outliers are on the left tail. Under both scenarios, the densities corresponding to the MDPDEs fit the bulk of the data better compared to the fitted densities via the MLEs. Due to assigning more probability to the outliers, the MLEs lead to underestimation near the mode of the data for all four cases. Despite LME being theoretically more robust than MLEs in the presence of outliers, we see that for all four subdivision-month pairs, it performs almost equally compared to MLE. For the pair (Chattisgarh; July), the fitted density using LME is similar to that based on MDPDE at $\alpha = 0.1$. Similar to the case of the exponential distribution, the fitted densities corresponding to the MDPDE at $\alpha = 0.1$ have less fitting bias compared to the MLE ($\alpha = 0$) in the region of the bulk of the data. The fitted densities corresponding to the MDPDEs at $\alpha = 0.5$ and $\alpha = 1$ appear to be very similar for the second and third cases. Further, from the third through sixth columns of Table \ref{table_gamma_bestfits}, we see that the estimates of both the shape and the rate parameters increase with $\alpha$ for all the cases. The estimates based on LME are similar to those based on MLE or MDPDE at $\alpha=0.1$, which indicates that LME can provide robustness only moderately. Standard errors of the MDPDE estimates increase with $\alpha$, with steep changes observable when moving from $\alpha = 0.1$ to $\alpha = 0.5$ for the fourth combination. The standard errors based on LME are again similar to those based on MLE or MDPDE at $\alpha=0.1$. The WDs based on MDPDE with $\alpha > 0$ are smaller than the WDs based on both MLE and LME for all four cases. In the fourth case, the WD for LME is larger than the WD based on MLE. Theoretically, under the pure data scenario explained in Section \ref{subsec:are}, the SEs of the MDPDEs are larger than the SEs of the MLEs. However, under practical settings with possible noise in the data, there is no guaranteed ordering between the SEs of MLE and MDPDE as seen in Table \ref{table_gamma_bestfits}.
Rather, MLE is known to have higher SE than the MDPDEs with a high tuning parameter choice. This ordering has been observed repeatedly under different settings within the vast literature of the MDPDE and robust statistics in general.

The WDs based on $\alpha = \alpha^\ast$ for all four cases are significantly smaller than the respective WDs based on LME. These MDPDEs at $\alpha = \alpha^\ast$ are larger than the respective MLEs as well as LMEs. The standard deviation (SD) of a gamma($a, b$) distributed random variable is $\sqrt{a}/b$; based on the MLEs, the SDs for the four cases are 0.3441, 0.2131, 0.3229, and 0.5298, respectively. The respective SDs based on LME are 0.3447, 0.1975, 0.3089, and 0.5413, while based on the MDPDEs at $\alpha = \alpha^\ast$, these SDs reduce to 0.2989, 0.1864, 0.2201, and 0.4148, respectively. These illustrate that both MLE and LME overestimate the model variance due to the presence of outliers, which can be solved successfully through our MDPDE approach. For all four subdivision-month combinations, $\alpha^\ast$ values are moderate and the corresponding $\hat{a}$, $\hat{b}$, $\textrm{SE}(\hat{a})$, and $\textrm{SE}(\hat{b})$ are also in between their values based on the extreme choices $\alpha=0$ and $\alpha=1$ in all cases indicating a balance in the bias-variance trade-off.}

% The corresponding optimal MDPDE estimates of the (shape, rate) parameters are (9.5238, 0.0198), (29.6784, 0.0778), (15.6104, 0.0440), and (6.5393, 0.1073), respectively. 

% Therefore, once again we expect to obtain an optimal $\alpha$ between 0.1 and 1 that would lead to the `best' fit for the data via appropriate trade-offs. 

% Based on the minimization of the WDs, the optimal values of $\alpha$ for these four subdivision-month pairs turn out to be 0.2562, 0.2303, 0.1277, and 0.1798, respectively, with the corresponding WDs being 0.0230, 0.0151, 0.0146, and 0.0356, respectively.

\subsection{{\color{black}MDPDE with lognormal distribution}}

\begin{figure}[h]
    \centering
\adjincludegraphics[width = 0.45\linewidth, trim = {{.0\width} {.0\width} {.0\width} {.0\width}}, clip]{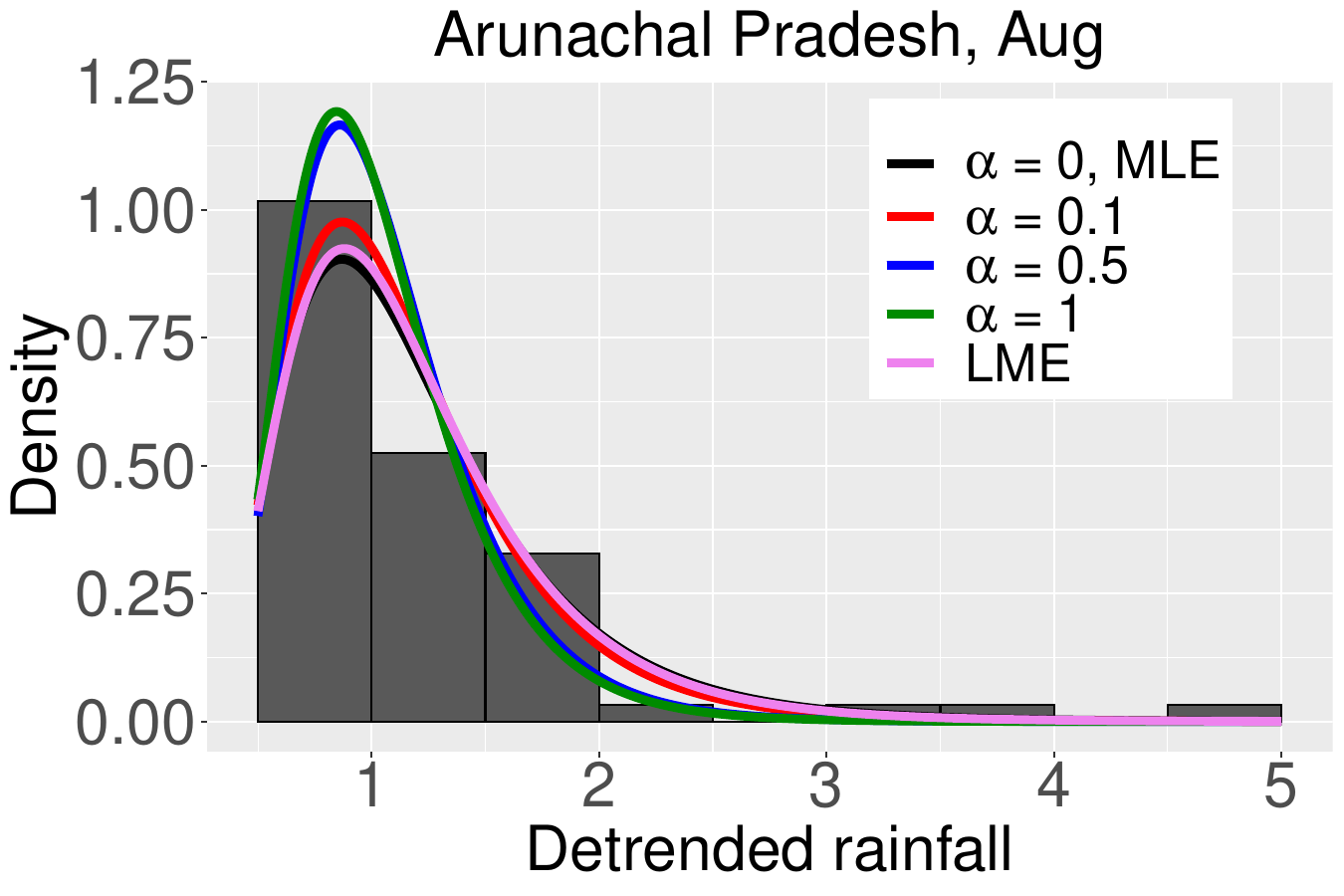}
\adjincludegraphics[width = 0.45\linewidth, trim = {{.0\width} {.0\width} {.0\width} {.0\width}}, clip]{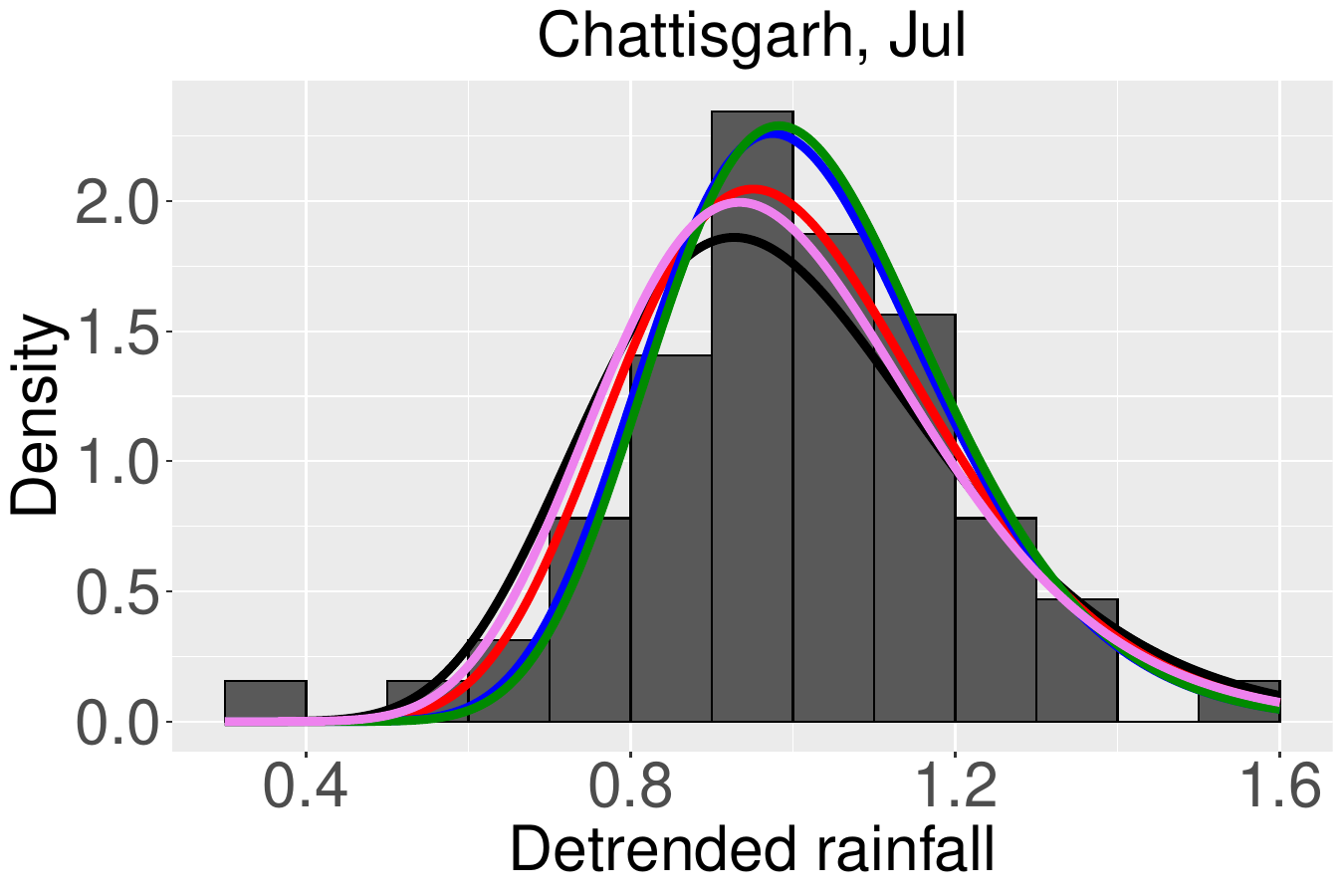} \\
\adjincludegraphics[width = 0.45\linewidth, trim = {{.0\width} {.0\width} {.0\width} {.0\width}}, clip]{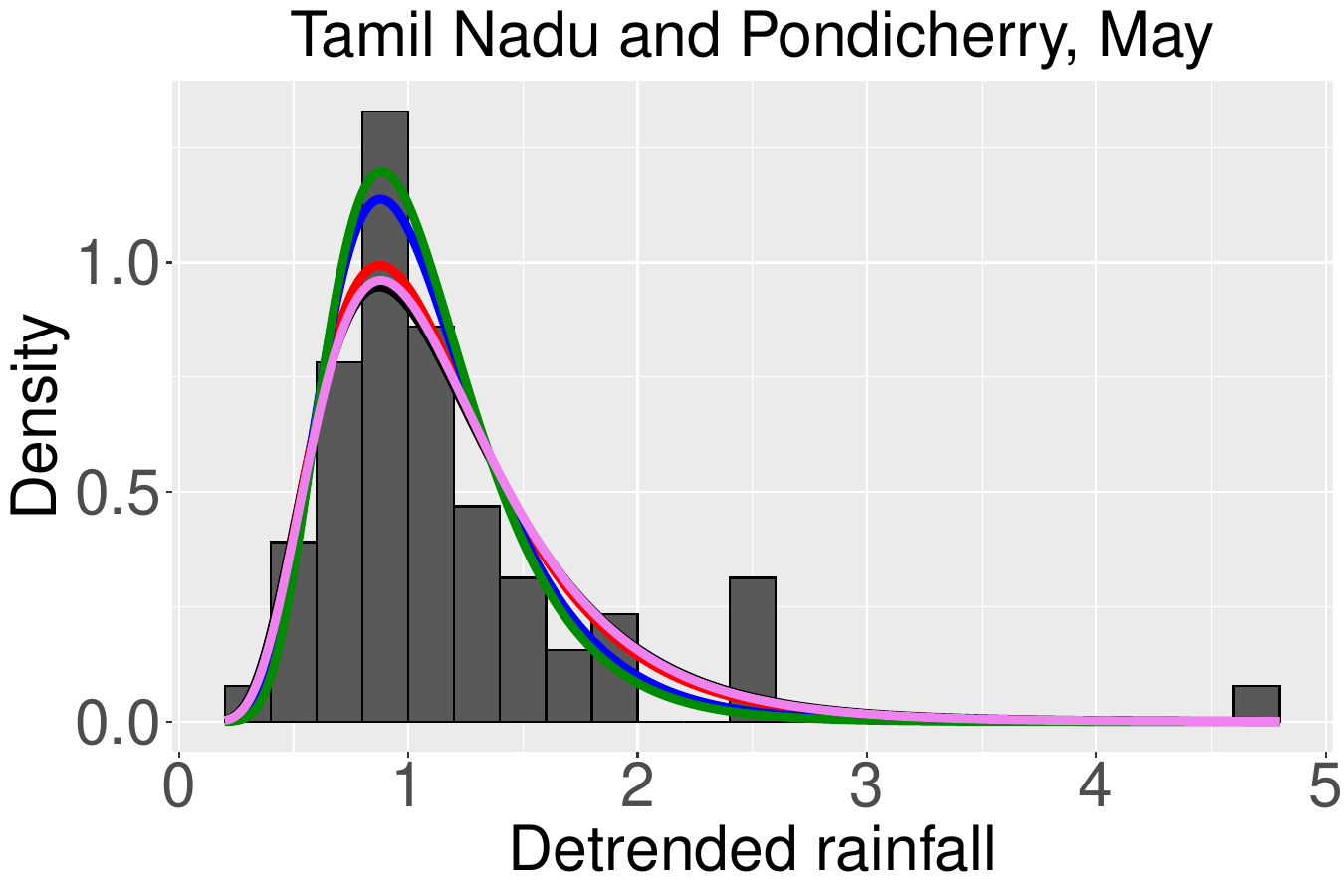}
\adjincludegraphics[width = 0.45\linewidth, trim = {{.0\width} {.0\width} {.0\width} {.0\width}}, clip]{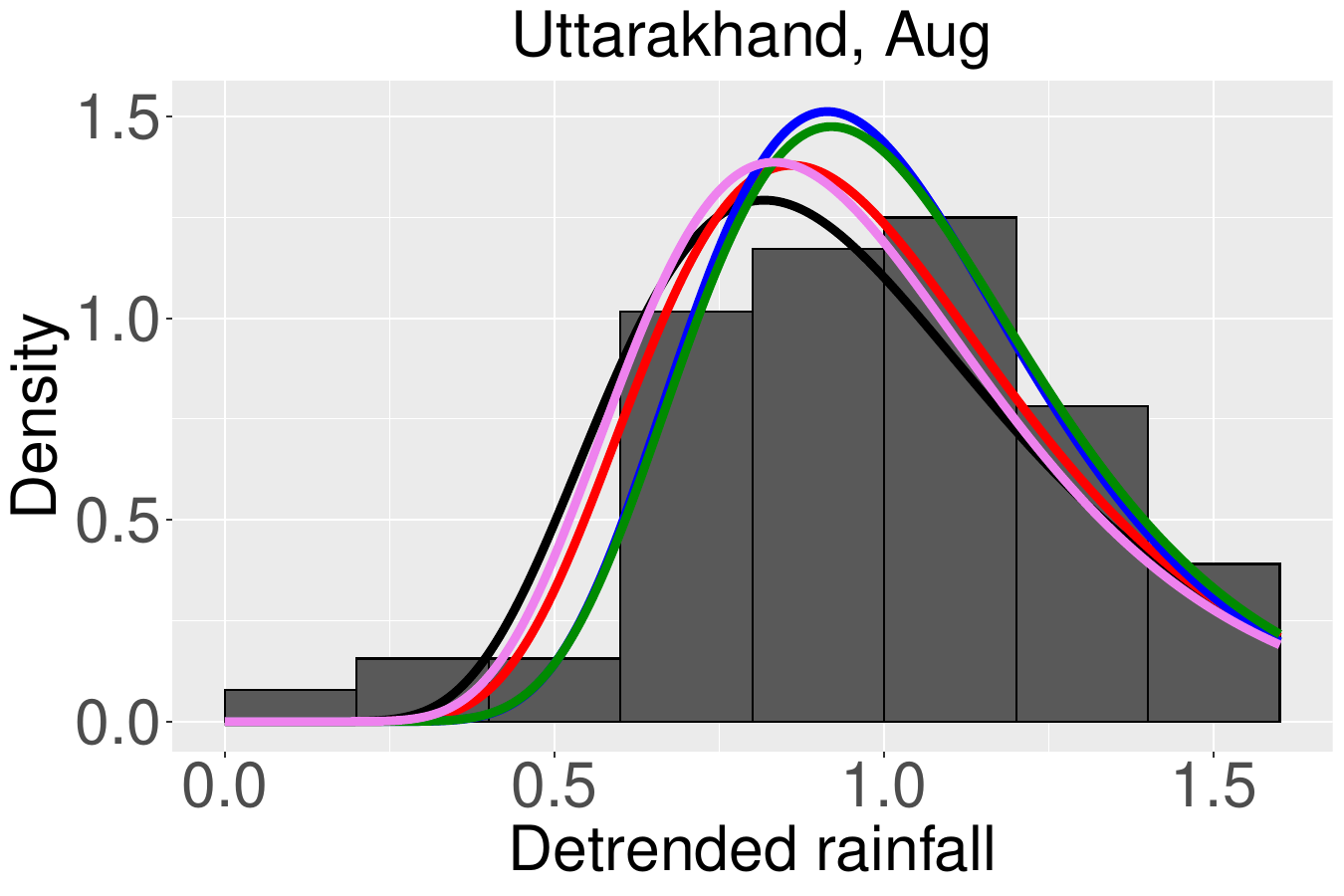}
    \caption{Histograms along with fitted lognormal densities {\color{black} to detrended rainfall data} with parameters estimated using MDPDE at $\alpha = 0, 0.1, 0.5, 1$, and LME. The MLE results coincide with the MDPDE case at $\alpha=0$. All the sub-figures share the same legend as in the top-left panel.}
    \label{fig_lognormal_bestfits}
\end{figure}

We further pick four subdivision-month combinations where lognormal distribution provides reliable fits and the data include outliers. The chosen cases are -- (Arunachal Pradesh; August), (Chattisgarh; July), (Tamilnadu and Pondicherry; May), and (Uttarakhand; August), with the proportion of outliers being {\color{black}4.9\%, 3.1\%, 9.4\%, and 3.1\%}, respectively.
The fitted lognormal density functions obtained using the MDPDEs and LMEs are provided in Figure \ref{fig_lognormal_bestfits}; the MDPDE and LME values, their SEs, and the associated WDs are provided in Table \ref{table_lognormal_bestfits}.

\begin{table}[h]
\caption{{\color{black}Parameter estimates, corresponding standard errors, and empirical Wasserstein distances (WD) for fitting lognormal distributions to detrended rainfall data at selective subdivision-month pairs, using MDPDE, MLE, and LME. The values of $\alpha^\ast$ for the four subdivision-month combinations are 0.2485, 0.2251, 0.2737, and 0.1972, respectively.}}
\label{table_lognormal_bestfits}
\centering
{\color{black}
\begin{tabular}{cccccccc}
\hline
\begin{tabular}[c]{@{}c@{}}Region, \\ Month\end{tabular} & Results  & $\alpha = 0$/MLE    & $\alpha = 0.1$  & $\alpha = 0.5$  & $\alpha = 1$  & LME & $\alpha = \alpha^\ast$.  \\
\hline     
Arunachal   & $\hat{\mu}$ & 0.0698 & 0.0447 & -0.0129 & -0.0309 & 0.0698 & 0.0125 \\ 
Pradesh,   & $\hat{\sigma}$ & 0.4573 & 0.4284 & 0.3714 & 0.3696 & 0.4443 & 0.3916 \\ 
August   & $\textrm{SE}(\hat{\mu})$ & 0.0577 & 0.0565 & 0.0561 & 0.0612 & 0.0577 & 0.0554 \\ 
   & $\textrm{SE}(\hat{\sigma})$ & 0.0531 & 0.0500 & 0.0428 & 0.0487 & 0.0500 & 0.0453 \\ 
   & WD  & 0.0383 & 0.0280 & 0.0303 & 0.0417 & 0.0375 & 0.0231 \\
   \hline    
Chattisgarh,  & $\hat{\mu}$ & -0.0244 & -0.0098 & 0.0063 & 0.0131 & -0.0244 & -0.0011 \\ 
July  & $\hat{\sigma}$ & 0.2254 & 0.2009 & 0.1783 & 0.1747 & 0.2094 & 0.1855 \\ 
   & $\textrm{SE}(\hat{\mu})$ & 0.0283 & 0.0252 & 0.0235 & 0.0258 & 0.0283 & 0.0234 \\ 
   & $\textrm{SE}(\hat{\sigma})$ & 0.0364 & 0.0294 & 0.0228 & 0.0247 & 0.0281 & 0.0239 \\ 
   & WD  & 0.0420 & 0.0237 & 0.0212 & 0.0305 & 0.0363 & 0.0165 \\ 
   \hline    
Tamilnadu  & $\hat{\mu}$ & 0.0572 & 0.0433 & 0.0093 & 0.0010 & 0.0572 & 0.0240 \\ 
\& Pondicherry,  & $\hat{\sigma}$ & 0.4382 & 0.4199 & 0.3725 & 0.3545 & 0.4301 & 0.3938 \\ 
May  & $\textrm{SE}(\hat{\mu})$ & 0.0538 & 0.0520 & 0.0504 & 0.0509 & 0.0538 & 0.0506 \\ 
  & $\textrm{SE}(\hat{\sigma})$ & 0.0475 & 0.0447 & 0.0487 & 0.0530 & 0.0458 & 0.0453 \\ 
   & WD  & 0.0318 & 0.0265 & 0.0243 & 0.0286 & 0.0306 & 0.0209 \\ 
   \hline    
Uttarakhand, & $\hat{\mu}$ & -0.0754 & -0.0487 & -0.0132 & -0.0036 & -0.0754 & -0.0320 \\ 
August & $\hat{\sigma}$ & 0.3543 & 0.3196 & 0.2778 & 0.2825 & 0.3273 & 0.2943 \\ 
  & $\textrm{SE}(\hat{\mu})$ & 0.0444 & 0.0404 & 0.0382 & 0.0432 & 0.0444 & 0.0384 \\ 
  & $\textrm{SE}(\hat{\sigma})$ & 0.0578 & 0.0498 & 0.0303 & 0.0323 & 0.0443 & 0.0413 \\ 
   & WD & 0.0460 & 0.0276 & 0.0286 & 0.0362 & 0.0404 & 0.0221 \\ 
   \hline
\end{tabular}
}
\end{table}

% In the fourth case, there is only one outlier (1.6\%) in the data corresponding to the first bin of the histogram.
%  even in a single outlier
%  they increase with $\alpha$ for the other two cases having outliers on the left tail. 
% , and these MDPDEs at $\alpha = \alpha^\ast$ are larger than the respective MLEs as well as LMEs

{\color{black} 
Figure \ref{fig_lognormal_bestfits} shows that the outliers are on the right tail for the first and the third cases while they are on the left tail for the other two cases. Considering both scenarios, the densities based on MDPDEs fit the bulk of the data better than those based on MLEs. For the first and third cases, the fitted densities based on MLE and LME are close to each other, while for the other two cases, the fitted densities using LME are similar to those based on MDPDE at $\alpha = 0.1$. Further, from Table \ref{table_lognormal_bestfits} we observe that the estimates of both $\mu$ and $\sigma$ decrease with $\alpha$ for the first and the third cases where outliers are on the right tail. For the other two cases, the estimates of $\mu$ increases with $\alpha$ while the estimates of $\sigma$ generally show the opposite trend. The SEs of the MDPDE estimates change only slightly with changing $\alpha$. The WDs corresponding to $\alpha = 0.1$ and $\alpha = 0.5$ are smaller than those for $\alpha = 0$ and $\alpha = 1$. The MLEs and LMEs of $\mu$ coincide theoretically (after transforming the data into the log scale), and for $\sigma$, LMEs are slightly smaller than the MLEs. Despite the WDs corresponding to LMEs being smaller than those based on MLEs, the MDPDEs with $\alpha=0.1$ and $\alpha=0.5$ provide even smaller WDs compared to LMEs for all the cases.

Similar to the case of fitting gamma distribution, the WDs based on $\alpha = \alpha^\ast$ values for all four cases are smaller than the respective WDs based on LME. The MDPDEs of $\mu$ at $\alpha=\alpha^\ast$ are smaller than the respective MLEs and LMEs for the first and the third cases, whereas the reverse order is observed for the rest. For $\sigma$, the MDPDEs at $\alpha=\alpha^\ast$ are smaller than the respective MLEs and LMEs for all four cases. The LME method provides smaller estimates of $\sigma$ compared to MLE for all four cases. As $(\mu, \sigma)$ denote the mean and standard deviation in the log scale, the patterns in Table \ref{table_lognormal_bestfits} indicate that the MDPDE method moves the fitted density towards the bulk of the data by removing the erroneous effects of the outliers.
}

% Based on previous inspections, we expect the `optimum' $\alpha$ to be between 0.1 and 0.5 for all four cases. By minimizing WDs, the optimal $\alpha$ for these four subdivision-month pairs indeed turn out to be 0.2546, 0.2448, 0.2194, and 0.1692, respectively, with the corresponding WDs being 0.0237, 0.0178, 0.0222, and 0.0203, respectively. The corresponding MDPDE estimates of $(\mu, \sigma)$ are (5.8625, 0.3899), (5.9323, 0.1829), (4.0978, 0.4080), and (5.9070, 0.3001), respectively. 
% The WDs corresponding to MDPDE with optimal $\alpha$'s are substantially smaller than those based on LME, which indicates that MDPDE provides better robustness than LME in the presence of outliers.

%\vspace{-2mm}
\subsection{{\color{black}MDPDE with Weibull distribution}}

\begin{figure}[h]
    \centering
\adjincludegraphics[width = 0.45\linewidth, trim = {{.0\width} {.0\width} {.0\width} {.0\width}}, clip]{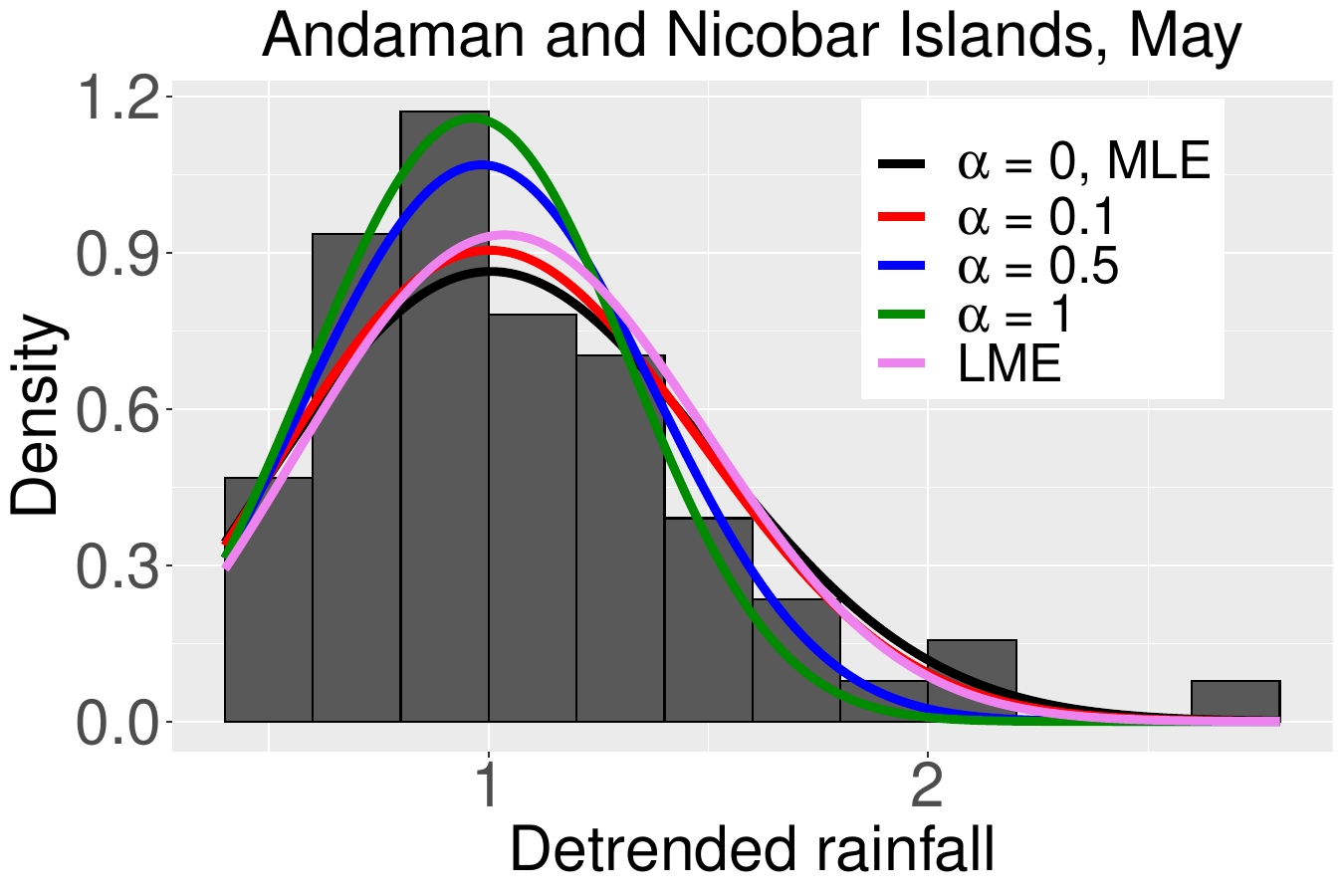}
\adjincludegraphics[width = 0.45\linewidth, trim = {{.0\width} {.0\width} {.0\width} {.0\width}}, clip]{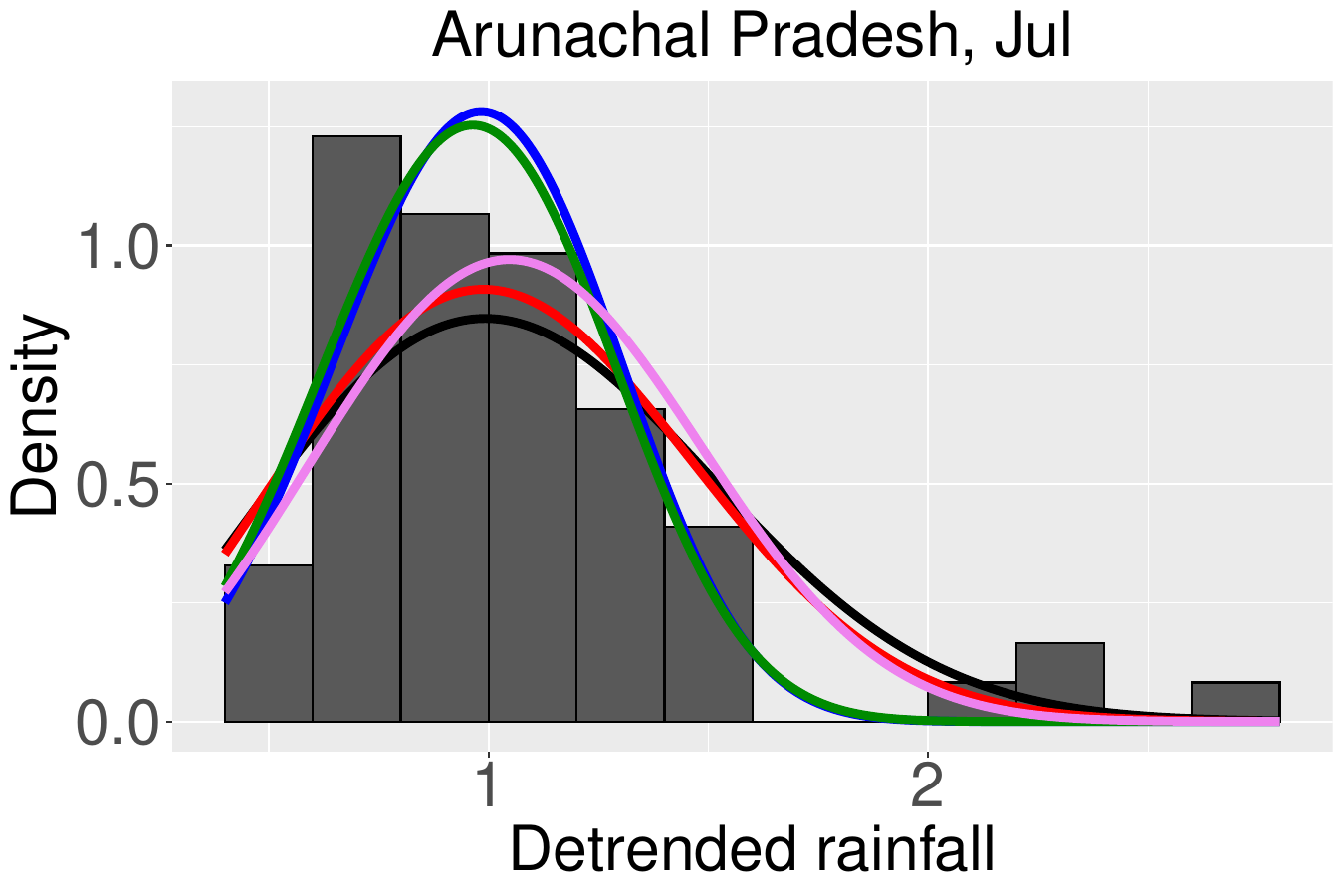} \\
\adjincludegraphics[width = 0.45\linewidth, trim = {{.0\width} {.0\width} {.0\width} {.0\width}}, clip]{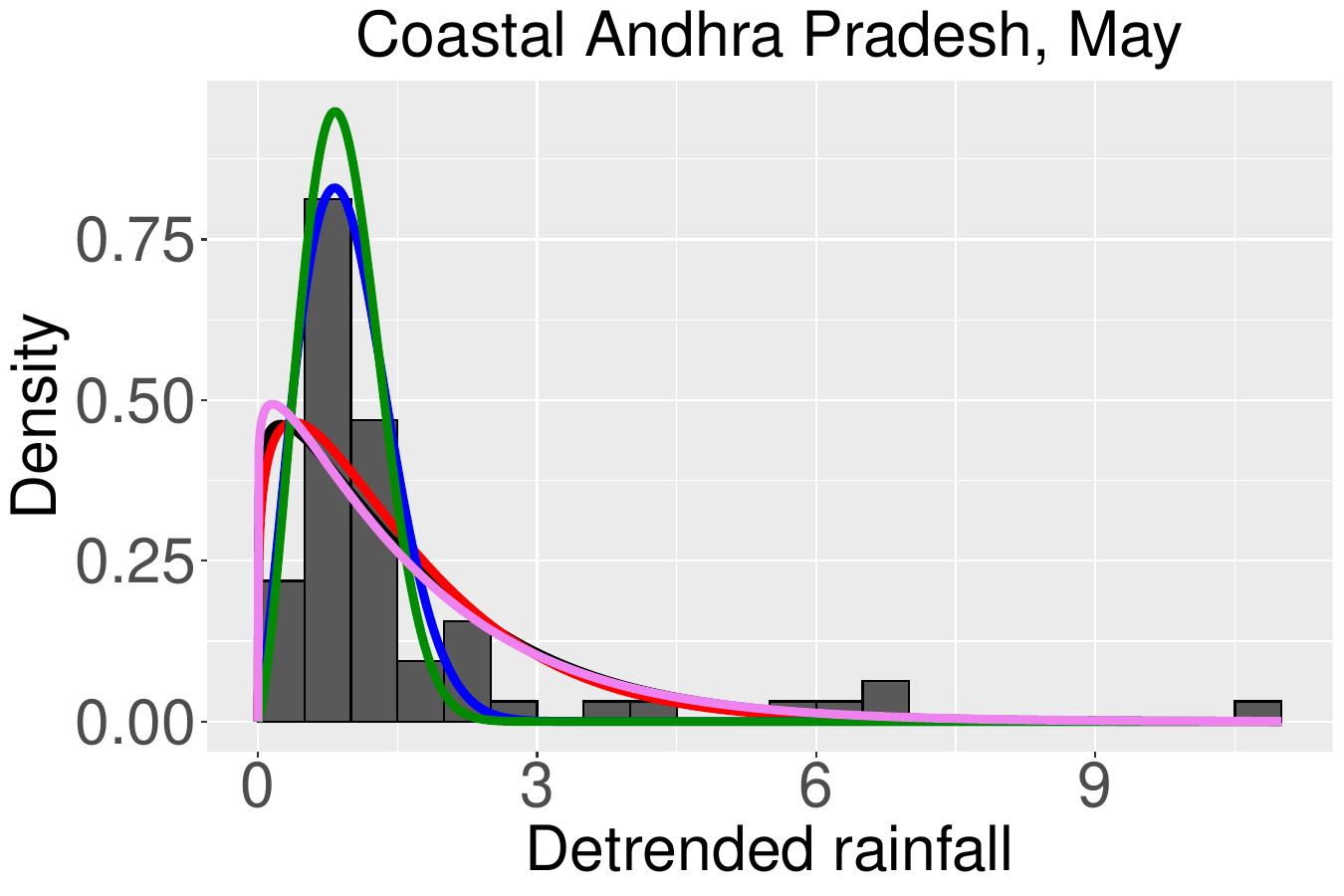}
\adjincludegraphics[width = 0.45\linewidth, trim = {{.0\width} {.0\width} {.0\width} {.0\width}}, clip]{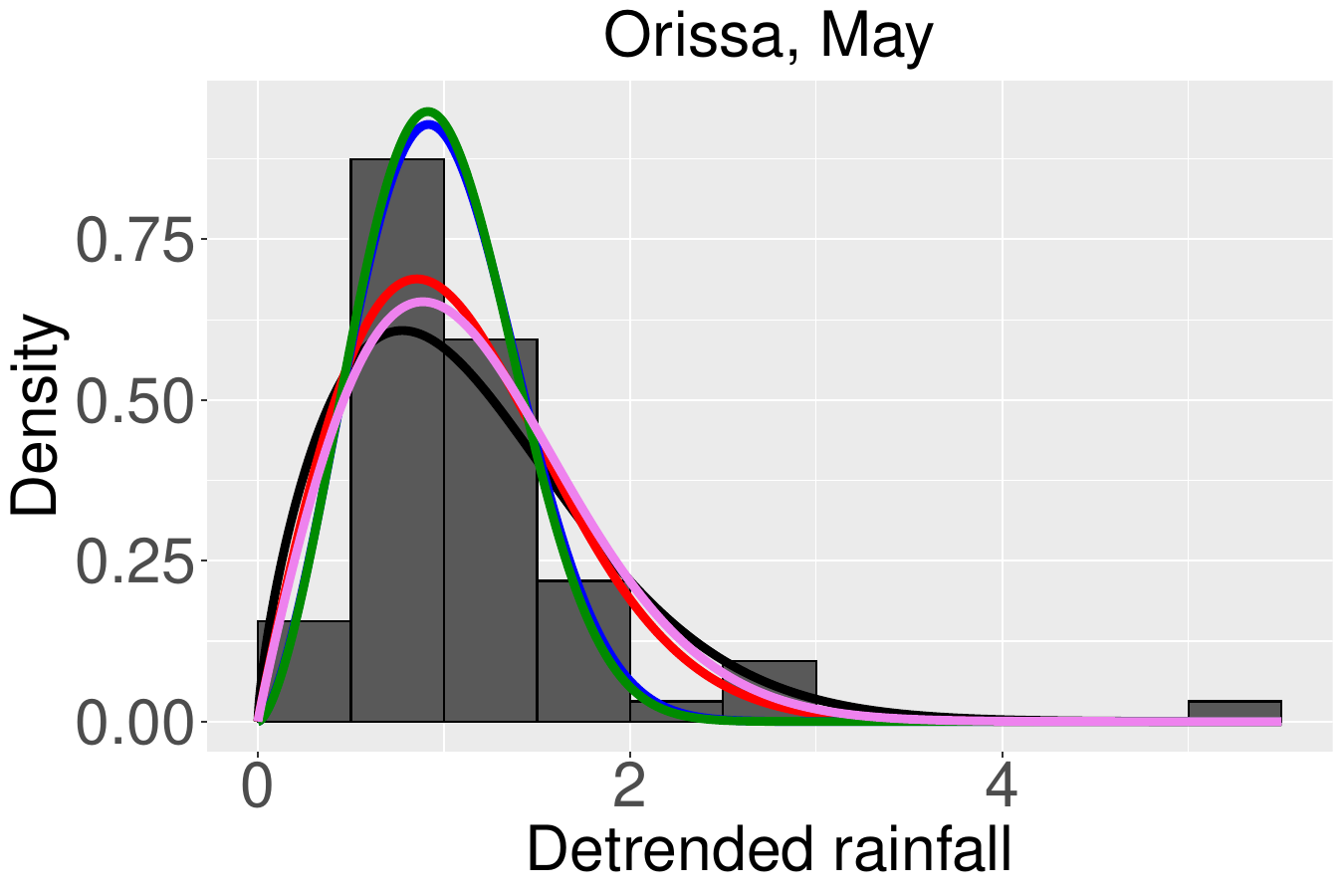}
    \caption{Histograms along with fitted Weibull densities {\color{black} to detrended rainfall data} with parameters estimated using MDPDE at $\alpha = 0, 0.1, 0.5, 1$, and LME. The MLE results coincide with the MDPDE case at $\alpha=0$. All the sub-figures share the same legend as in the top-left panel}
    \label{fig_weibull_bestfits}
\end{figure}

Finally, we choose four subdivision-month combinations where Weibull distribution provides reliable fits and the data include some outliers. The selected cases are -- (Andaman and Nicobar Islands; May), (Arunachal Pradesh; July), 
(Coastal Andhra Pradesh; May) and (Orissa; May) having outlier proportions as {\color{black} 4.7\%, 4.9\%, 17.2\%, and 6.2\%}, respectively. The fitted Weibull densities based on MDPDE and LME are provided in Figure \ref{fig_weibull_bestfits}, and the individual MDPDEs and LMEs along with their SEs and the corresponding WDs are provided in Table \ref{table_weibull_bestfits}.

% In each of the first and fourth cases, there is only one outlier (1.6\%) present in the data.

\begin{table}[ht]
\caption{{\color{black}Parameter estimates, corresponding standard errors, and empirical Wasserstein distances (WD) for fitting Weibull distributions to detrended rainfall data at selective subdivision-month pairs, using MDPDE, MLE, and LME. The values of $\alpha^\ast$ for the four subdivision-month combinations are 0.3591, 0.2306, 0.2782, and 0.2939, respectively.}}
\label{table_weibull_bestfits}
\centering
{\color{black}
\begin{tabular}{cccccccc}
\hline
\begin{tabular}[c]{@{}c@{}}Region, \\ Month\end{tabular} & Results  & $\alpha = 0$/MLE    & $\alpha = 0.1$  & $\alpha = 0.5$  & $\alpha = 1$ & LME & $\alpha = \alpha^\ast$  \\
\hline
Andaman   & $\hat{a}$ & 2.6103 & 2.7030 & 3.0644 & 3.2262 & 2.8570 & 2.9774 \\ 
\& Nicobar,  & $\hat{b}$ & 0.8269 & 0.8415 & 0.8933 & 0.9255 & 0.8295 & 0.8788 \\ 
May  & $\textrm{SE}(\hat{a})$ & 0.3025 & 0.3002 & 0.4236 & 0.5066 & 0.3270 & 0.3757 \\ 
  & $\textrm{SE}(\hat{b})$ & 0.0432 & 0.0443 & 0.0508 & 0.0553 & 0.0432 & 0.0488 \\ 
 & WD & 0.0399 & 0.0341 & 0.0292 & 0.0504 & 0.0382 & 0.0260 \\ 
\hline 
Arunachal   & $\hat{a}$ & 2.5417 & 2.6796 & 3.5922 & 3.4582 & 2.9738 & 3.1438 \\ 
Pradesh,   & $\hat{b}$ & 0.8276 & 0.8500 & 0.9288 & 0.9399 & 0.8319 & 0.8952 \\ 
July    & $\textrm{SE}(\hat{a})$ & 0.3620 & 0.4253 & 0.4260 & 0.4450 & 0.4129 & 0.5971 \\ 
  & $\textrm{SE}(\hat{b})$ & 0.0471 & 0.0495 & 0.0453 & 0.0542 & 0.0470 & 0.0537 \\ 
 & WD & 0.0598 & 0.0475 & 0.0425 & 0.0526 & 0.0549 & 0.0276 \\ 
\hline  
Coastal  & $\hat{a}$ & 1.1260 & 1.2151 & 2.1758 & 2.4139 & 1.0840 & 1.9027 \\ 
Andhra  & $\hat{b}$ & 0.5868 & 0.6210 & 0.9134 & 0.9658 & 0.5999 & 0.8419 \\ 
Pradesh,  & $\textrm{SE}(\hat{a})$ & 0.1154 & 0.1649 & 0.4826 & 0.4592 & 0.1697 & 0.4294 \\ 
May   &  $\textrm{SE}(\hat{b})$ & 0.0749 & 0.0882 & 0.0931 & 0.0760 & 0.0700 & 0.1224 \\ 
   & WD & 0.0871 & 0.0788 & 0.0714 & 0.0916 & 0.0863 & 0.0541 \\
\hline    
Orissa,   & $\hat{a}$ & 1.7160 & 1.9603 & 2.5680 & 2.6037 & 1.9394 & 2.3560 \\ 
May   & $\hat{b}$ & 0.7745 & 0.8129 & 0.9000 & 0.9092 & 0.7763 & 0.8696 \\ 
  & $\textrm{SE}(\hat{a})$ & 0.3080 & 0.3178 & 0.3614 & 0.3796 & 0.3153 & 0.3408 \\ 
 & $\textrm{SE}(\hat{b})$ & 0.0605 & 0.0614 & 0.0557 & 0.0558 & 0.0604 & 0.0582 \\ 
  & WD & 0.0573 & 0.0400 & 0.0332 & 0.0386 & 0.0486 & 0.0235 \\ 
   \hline
\end{tabular}
}
\end{table}

{\color{black}
From Figure \ref{fig_weibull_bestfits}, we observe that the densities based on MLEs have a similar trend of underestimation near the bulk and overestimation near the tails like the other three rainfall models. For $\alpha = 0.1$, the fitted densities have less bias compared to $\alpha = 0$ but have a similar pattern of bias. For $\alpha = 0.5$ and $\alpha = 1$, the fitted densities are approximately similar, specifically for the second and fourth examples. The fitted densities based on LME appear to fit the bulk of the data slightly more accurately than MLE indicating its effectiveness in terms of robustness; however, the fitted densities based on MDPDE with $\alpha > 0.1$ provide a better fit near the mode of the empirical distribution compared to those based on LME. From Table \ref{table_weibull_bestfits}, we observe that the estimates of the parameters $a$ and $b$ increase as we move from $\alpha = 0$ to $\alpha = 0.1$ for all the cases. A further increase in $\alpha$ increases the estimates of $b$ for all the cases and the estimates of $a$ also show a similar pattern in most of the cases. For most of the cases, standard errors of the MDPDEs (specifically for $b$) change only moderately with changing $\alpha$. The WDs corresponding to $\alpha = 0.1$ and $\alpha = 0.5$ are again significantly smaller compared to those for $\alpha = 0$ and $\alpha = 1$ as well as LME for almost all cases. 

Similar to the example cases for the lognormal distribution, we again expect $\alpha^\ast$ values to be between 0.1 and 0.5. All four WD values are substantially smaller than the respective WDs based on MLE and LME. The MDPDE estimates for $\alpha = \alpha^\ast$ are larger than the respective MLEs as well as LMEs. The mean and SD of a Weibull($a, b$) distributed random variable are $b^{-1}\Gamma(1 + 1/a)$ and $b^{-1}[\Gamma(1 + 2/a) - \Gamma^2(1 + 1/a)]^{1/2}$, respectively; based on MLEs, the means for the four cases are 1.0742, 1.0725, 1.6322, and 1.1513, respectively, while based on the MDPDEs with $\alpha = \alpha^\ast$, these means become 1.0158, 0.9996, 1.0540, and 1.0191, respectively. The respective means based on LME are 1.0743, 1.0730, 1.6164, and 1.1424, which are close to the means based on MLE. Thus, the (optimum) MDPDE shifts the fitted Weibull densities toward zero by removing the effect of the outliers on the right tail. Further, the SDs based on MLE for these four cases are 0.4422, 0.4522, 1.4523, and 0.6912, respectively, and the SDs based on LME for the respective cases are 0.4079, 0.3931, 1.4924, and 0.6139; except for the third case, LME provides less SD compared to MLE. Based on the MDPDEs at $\alpha=\alpha^\ast$ values, SDs reduce to 0.3717, 0.3484, 0.5763, and 0.4599, respectively. Thus, we see that MLE and LME both overestimate the model variance in the presence of outliers, which is corrected via the proposed approach using the optimum MDPDE.
}

% By minimizing the WD values, we get the optimal $\alpha$ for these four subdivision-month pairs as 0.3254, 0.2126, 0.2550, and 0.2362, respectively, with the corresponding WDs being 0.0288, 0.0288, 0.0463, and 0.0268, respectively. 

%The corresponding MDPDE estimates of the shape and the rate parameter pairs $(a, b)$ are (2.9649, 0.0026), (3.0736, 0.0017), (1.6102, 0.0209), and (2.2017, 0.0151), respectively, which are larger than the respective MLEs as well as LMEs. 

%\vspace{-2mm}
\subsection{Comprehensive model selection using the RIC}\label{subsec:final_selection}

Here we finalize the appropriate RM for each subdivision-month combination to obtain a comprehensive picture of Indian rainfall distribution {\color{black}(after a suitable detrending)}. We follow the MDPDE-based robust model selection with the RIC as described in Subsection \ref{model_selection} and the `best' selected models for all the cases are reported in the Supplementary Material.

{\color{black}
Out of the total 432 subdivision-month pairs, the four rainfall models, namely exponential, gamma, lognormal, and Weibull are selected for 50, 101, 171, and 110 cases, respectively. Considering month-wise analysis, the exponential distribution is mostly selected for June (8 cases) and the least for April (one case). The gamma distribution is selected mostly for February (13 cases) and least for December (5 cases). The lognormal distribution is selected mostly for April (20 cases) and least for March (10 cases). The Weibull distribution is selected mostly for December (12 cases) and least for February (6 cases). Considering subdivision-wise analysis, the exponential distribution is selected mostly for Kerala (4 months) and never for 9 subdivisions. The gamma distribution is selected mostly for West Madhya Pradesh and Telangana (6 months) and only once for 7 subdivisions. The lognormal distribution is mostly selected for Uttarakhand and Himachal Pradesh (9 months) and least selected for Telangana (one case). The Weibull distribution is selected mostly for Lakshadweep (7 months) and least for Uttarakhand and Himachal Pradesh (zero cases).} Thus, instead of considering a particular model which is often done in the literature, we discuss a method for model selection along with a robust estimation approach that provides better inference at a more granular level of meteorological subdivisions in India.

%\vspace{-2mm}
\subsection{{\color{black}Projected future} rainfall amounts}

Based on the models selected by RIC and the model parameters estimated by the proposed MDPDE approach with the optimal tuning parameter, we finally calculate the median rainfall amounts for each subdivision-month pair. {\color{black}During the detrending procedure mentioned in Section \ref{data}, we remove the temporal trend, a function of the year of interest, from the raw dataset (in log scale). As a result, the projected median rainfall amounts in the original scale depend on the specific choice of the year. For illustrative purposes, we provide the results for the year 2025 in the supplementary material. Similarly, the rainfall amounts for percentiles, other than the median and also for other years, can be calculated as per the requirements.}

Although our modeling is done considering only the months that received nonzero rainfall, there is a certain percentage of zero observations in the data corresponding to dry months. For such dry months, we adjust the fitted distributions to get the estimated median rainfall amount as follows. If the proportion of dry months is more than 50\%, the estimated amount is zero; otherwise, if the proportion of dry months is $100p$\%, the estimated amount is the $[(50 - 100p) / (1 - p)]$-th percentile of the fitted probability distribution. This is a common strategy for zero-inflated rainfall data modeling while considering the dry and wet periods jointly. 

\begin{figure}[h]
	\centering
	\adjincludegraphics[width = 0.45\linewidth, trim = {{.0\width} {.0\width} {.0\width} {.0\width}}, clip]{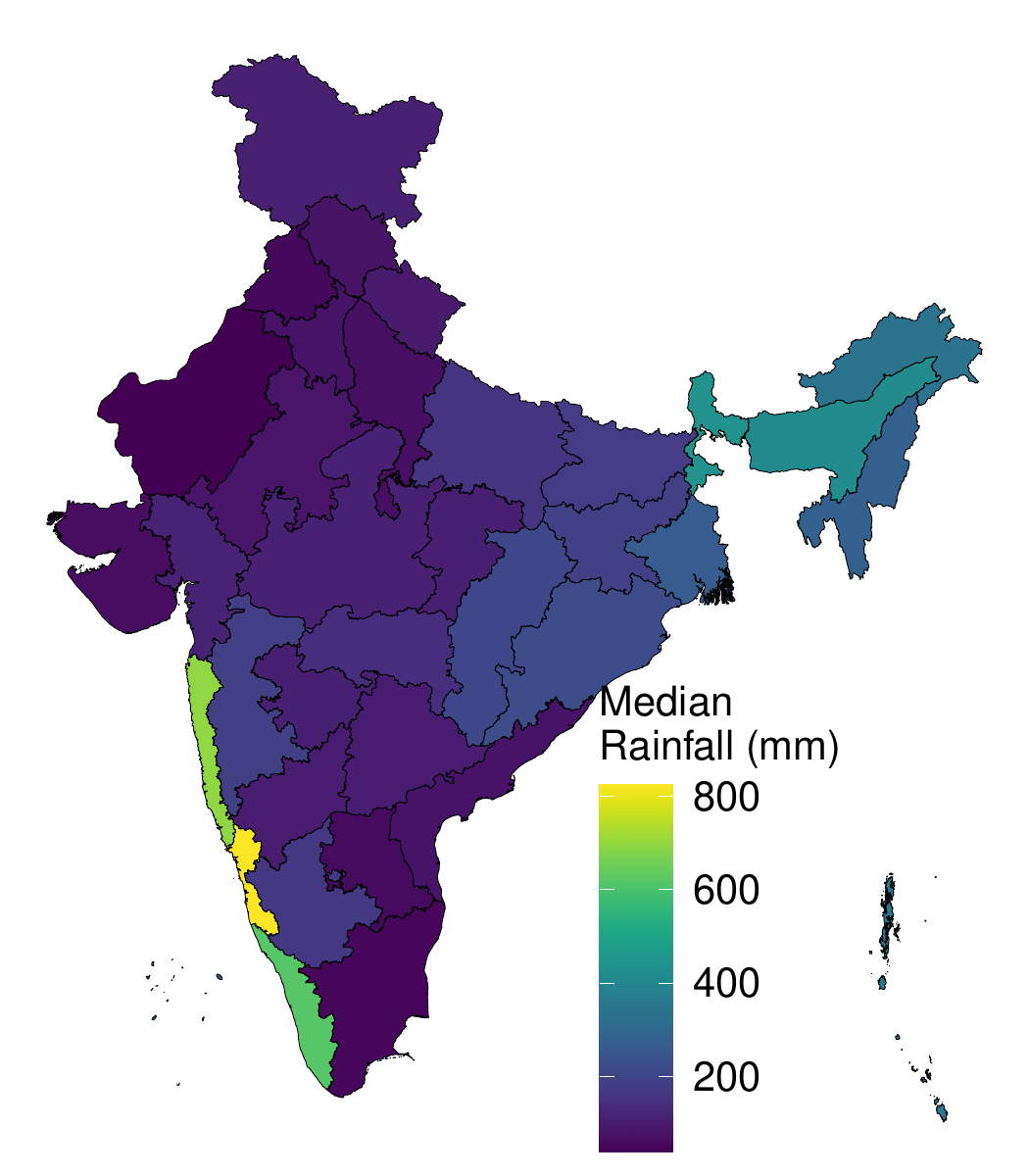}
\adjincludegraphics[width = 0.45\linewidth, trim = {{.0\width} {.0\width} {.0\width} {.0\width}}, clip]{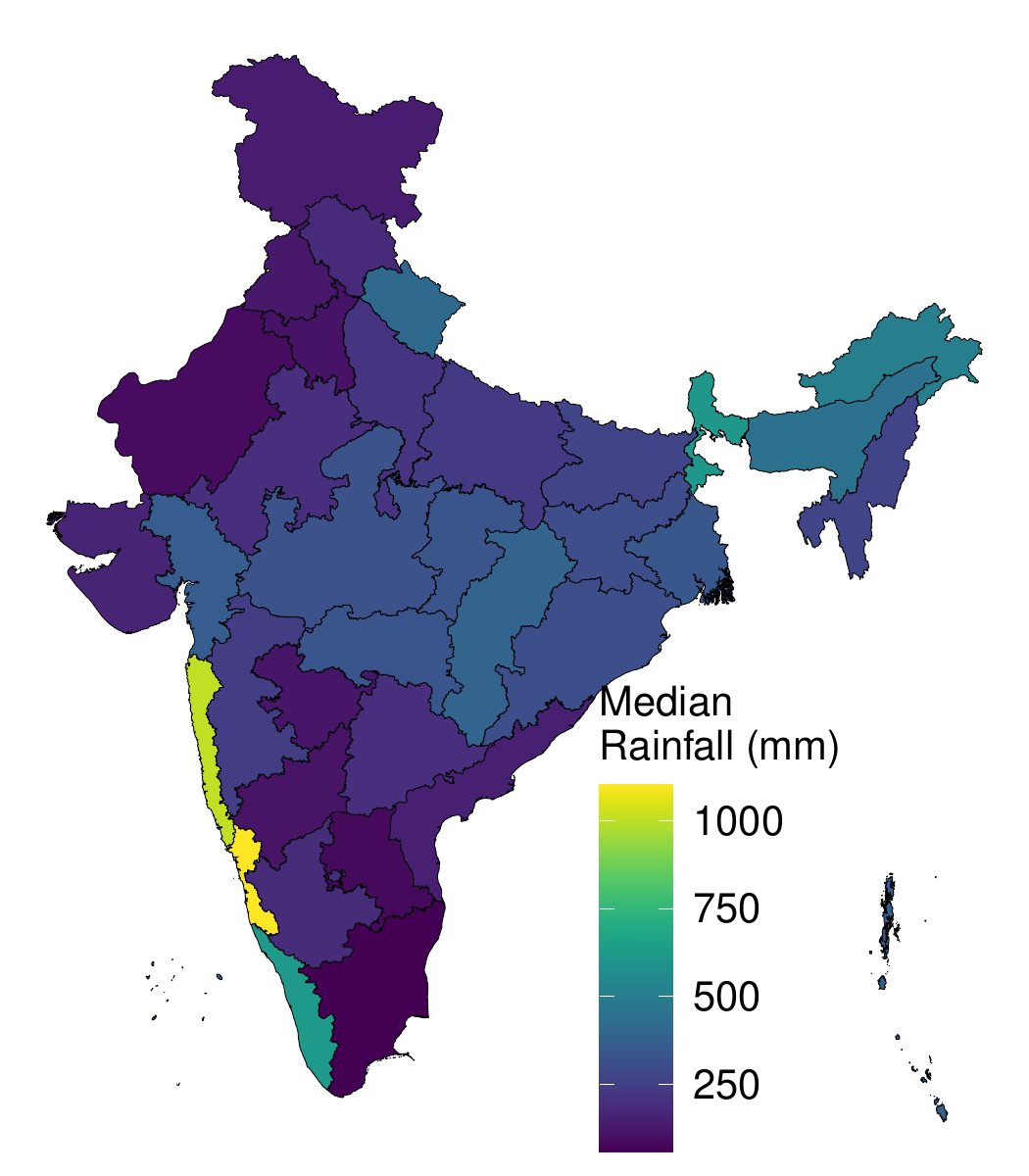}
\adjincludegraphics[width = 0.45\linewidth, trim = {{.0\width} {.0\width} {.0\width} {.0\width}}, clip]{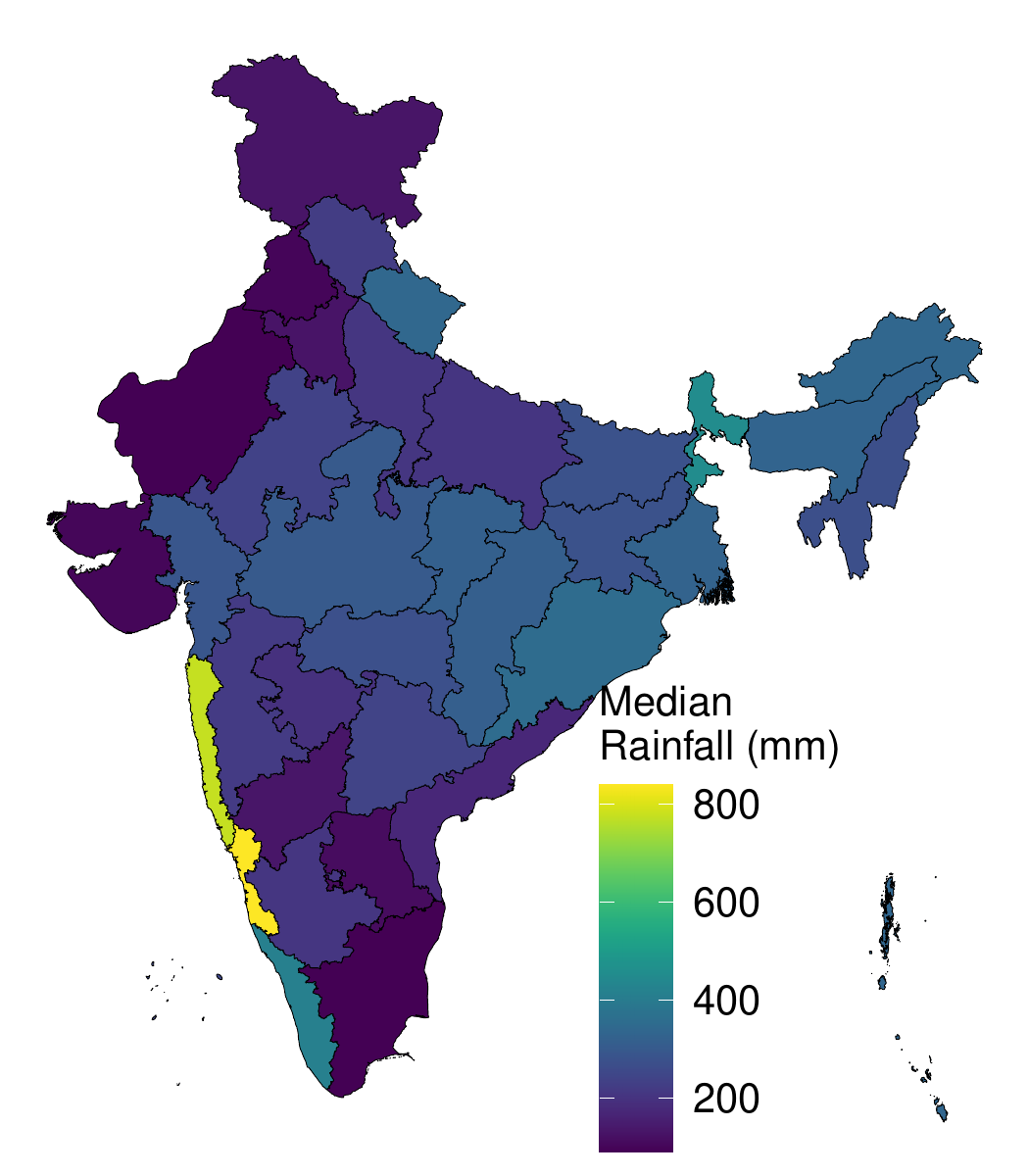}
\adjincludegraphics[width = 0.45\linewidth, trim = {{.0\width} {.0\width} {.0\width} {.0\width}}, clip]{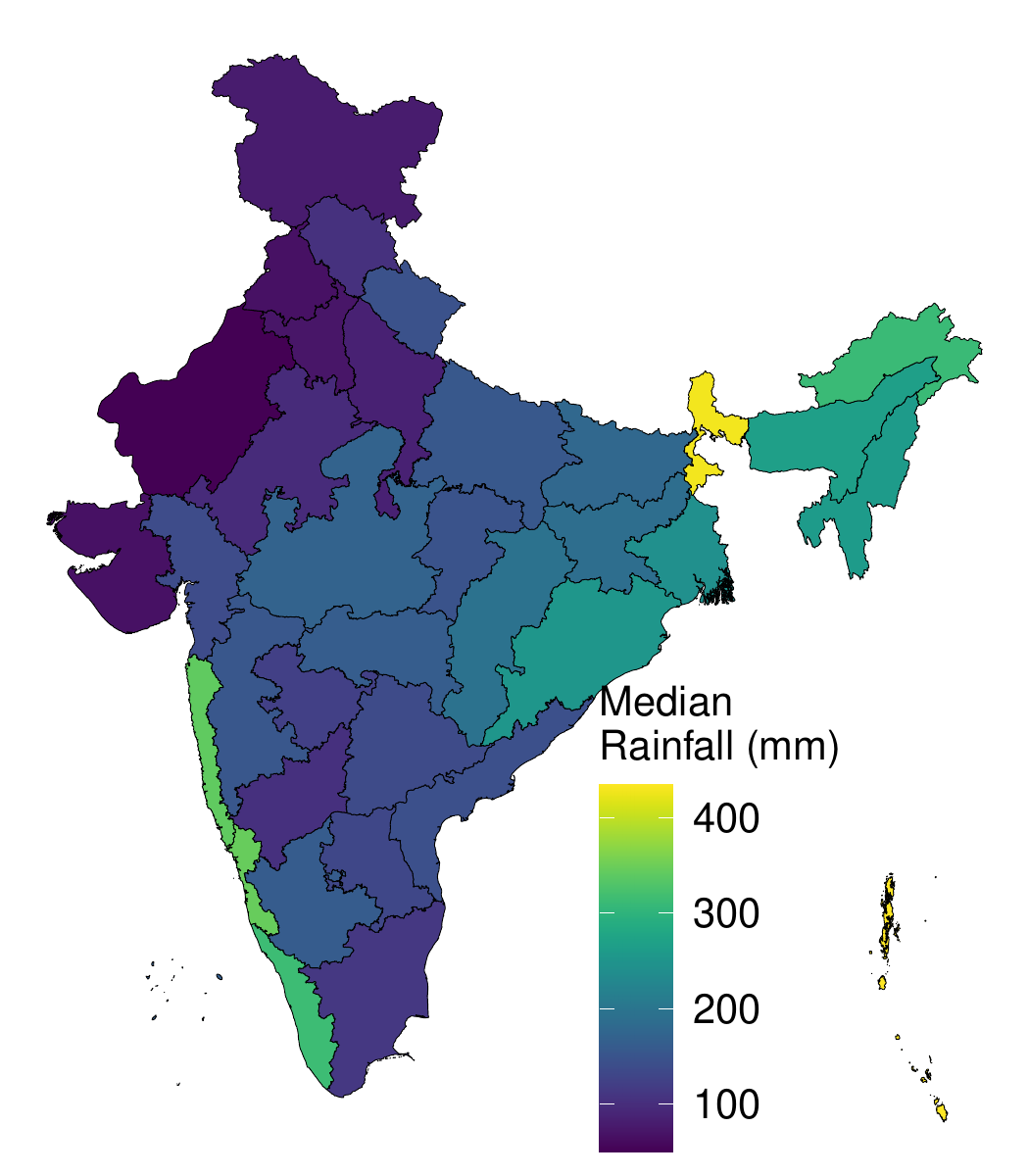}
\caption{Spatial maps of the estimated median rainfall amounts {\color{black}projected for the year 2025} for different meteorological subdivisions for June (top-left), July (top-right), August (bottom-left), and September (bottom-right)}
	\label{fig_spatialmaps_median}
\end{figure}

For June through September, the monsoon months, the amounts of areally-weighted rainfall are high across all the subdivisions of India. The estimated median rainfall amounts based on our methodology follow a similar seasonal pattern as well; the spatial maps of the estimated median rainfall amounts {\color{black}for the year 2025} and different meteorological subdivisions are presented in Figure \ref{fig_spatialmaps_median}. {\color{black}For the three months of June through August, the median rainfall amounts are maximum in the Coastal Karnataka subdivision, and in September, it is maximum for the subdivision Andaman and Nicobar Islands. In July, the median rainfall amount is minimum for the subdivision Tamil Nadu-Pondicherry, and during the other three monsoon months, it is minimum for the West Rajasthan subdivision. For the pre-monsoon month of May and the post-monsoon month of October, the median rainfall amounts are high in the Andaman and Nicobar Islands, north-eastern and southern subdivisions. In May, the projected amount of rainfall is highest (305.39 mm) for the subdivision Andaman and Nicobar Islands and lowest for Saurashtra, Kutch, and Diu (0.81 mm). In October, the amount is highest (309.82 mm) for the subdivision Kerala and lowest for the Haryana, Chandigarh, and Delhi subdivision (1.13 mm). For the other months, the amounts of rainfall are low except for the northeastern and northern sub-Himalayan subdivisions. The average monthly median rainfall is maximum in the Coastal Karnataka subdivision (287.80 mm) and minimum in the West Rajasthan subdivision (24.49 mm).}

% Out of 432 subdivision-month pairs, median rainfall amounts are zero for 14 cases. 

%\vspace{-4mm}

\section{Discussions and conclusions}
\label{conclusion}

The MLE is the most widely used parameter estimation procedure in the meteorological literature and other disciplines due to some theoretical properties and the availability of software for their computations. However, they are sensitive to outliers and are strongly affected even in the presence of a single outlier. The presence of outliers is a common issue in rainfall data, and hence, a robust parameter estimation approach is required to estimate the model parameters more accurately. Here, we discuss an easily implementable robust parameter estimation procedure, namely the MDPDE of \cite{basu1998robust}, where we obtain the estimates by minimizing a density-based divergence measure. While the applications of MDPDE are in diverse scientific areas, it has not been explored for modeling monthly or annual rainfall. The exponential, gamma, Weibull, and lognormal probability distributions are widely used for rainfall modeling. We study how MDPDE performs for these rainfall models and discuss choosing an optimal value of the underlying robustness tuning parameter. While AIC is used for model comparison if the model parameters are estimated by MLE, we discuss RIC as a model selection criterion for MDPDE. We provide codes written in \texttt{R} (\url{http://www.R-project.org/}) for the estimation of the parameters by the MDPDE, calculating their standard errors by bootstrapping, finding the optimal tuning parameter, and calculating RIC for model comparison.

Apart from discussing the statistical method of the MDPDE for robust parameter estimation in rainfall data, 
we analyze areally-weighted monthly rainfall data from the 36 meteorological subdivisions of India for the years 1951--2014, where a substantial amount of outliers are present in the data. We fit the four rainfall models and estimate the model parameters using the MDPDE for all subdivision-month combinations. 
For each rainfall model, we present results at four subdivision-month combinations to illustrate the advantage of the MDPDE-based approach over the MLE approach. We provide tables of the best-fit models and the median rainfall amounts {\color{black} for the year 2025} estimated based on the MDPDE from the best-fitted model (in the supplementary material). As per the report of NRAA, the rainfed agroecosystem is divided into five homogeneous production systems- i. The rainfed rice-based system, ii. The nutritious (coarse) cereals-based system, iii. The oil-seeds-based system, iv. Pulses-based system and v. Cotton-based system. Out of these, the rainfed rice-based system is most sensitive to the availability of water. Rainfed rice cultivation is prevalent in the northeastern through eastern (Chattisgarh) subdivisions. For the four subdivisions-- Gangetic West Bengal, Orissa, Jharkhand, and Chattisgarh, the rainfall amounts are low (below 100 mm on average) for May and October. Thus, proper irrigation facilities are necessary for long-duration cultivation or multiple cultivations within a year. Altogether, a risk assessment before sowing is crucial as the success of rainfed agriculture largely depends on the rainfall amounts. The estimated rainfall amounts and the availability of software for quantifying the associated risk would be highly beneficial for agricultural planners. % (less than 100 mm on average)

While the MDPDE was originally proposed by \cite{basu1998robust} for univariate data, \cite{chakraborty2020robust} recently discussed its applicability for multivariate data as well. Instead of treating the rainfall data across different regions as independent observations, multivariate/spatial/areal modeling has been widely considered in the spatial statistics literature \citep{cressie1993statistics} where we can obtain superior estimates by borrowing information from the neighboring regions. As of our knowledge, MDPDE has never been implemented in a spatial setting, and it is a possible future endeavor. In the context of approximate Bayesian inference for high-dimensional spatial data using two-stage models, the parameters of the marginal distributions are often estimated in a robust way \citep{johannesson2021approximate, hazra2021large} in the first stage, before fitting a spatial model in the second stage, and MDPDE can be readily used in such a scenario. In this paper, we focus on a small family of probability distributions mostly used for daily, monthly, or annual rainfall analysis, after adjusting for zero-inflation. However, some other reasonable choices include Pearson Type-V/VI, log-logistic, generalized exponential, and generalized gamma distributions. The theoretical exploration of MDPDE for such models is complex; while the generalized exponential case is explored by \cite{hazra2022minimum}, exploring the other cases would be a possible future endeavor. In the context of extreme rainfall analysis, MDPDE can be a possible alternative to MLE for estimating the parameters of the max-stable models like the generalized extreme value distribution and the generalized Pareto distribution and their generalizations like a four-parameter kappa distribution; these research directions are future endeavors.

%\backmatter
%\vspace{-4mm}

\section*{Acknowledgement}
The authors would like to thank an Associate Editor and two anonymous reviewers for their several thoughtful suggestions which improved the flow and the content of the paper substantially. The research of the first author is partially supported by the Indian Institute of Technology Kanpur and Rice University collaborative research grant under Award No. DOIR/2023246.

\section*{Supplementary information}

Codes (written in R) for calculating MDPDEs and their standard errors, optimal tuning parameter selection, and model selection are provided in the Supplementary Material (also available at \url{https://github.com/arnabstatswithR/robustrainfall.git}). Tables of the best-fit models and the median rainfall amounts estimated based on the MDPDE from the best-fitted models are also provided.
% for all subdivision-month pairs
%\vspace{-4mm}
% \section*{Acknowledgments}

% The authors would like to thank an Associate Editor and a Reviewer for their thoughtful suggestions which improved the quality of the paper substantially.

\section*{Disclosure statement}

No potential conflict of interest was reported by the authors.

\bibliography{sn-bibliography}% common bib file

\end{document}